\documentclass[preprint,sort&compress,3p]{elsarticle}

\pdfoutput=1

\usepackage{graphicx}
\usepackage{subfig}
\usepackage{miller}
\usepackage{amssymb}
\usepackage{amsthm}
\usepackage{amsmath}
\usepackage{array}
\usepackage{caption}
\usepackage[nodots]{numcompress}
\usepackage{booktabs}
\usepackage[protrusion=true,expansion=true]{microtype}
\usepackage[section]{placeins}
\DeclareMathOperator{\sgn}{sgn}
\usepackage{color}

\begin{document}

\begin{frontmatter}

\title{The candidacy of shuffle and shear during compound twinning in hexagonal close-packed structures}

\author[MSU,CAVS]{Haitham El Kadiri\corref{cor1}}
\cortext[cor1]{Corresponding author}
\ead{elkadiri@me.msstate.edu}
\author[MSU,CAVS]{Christopher D. Barrett}
\author[CAVS,ARO]{Mark A. Tschopp}
\address[MSU]{Department of Mechanical Engineering, Mississippi State University, Mississippi State, MS 39762, USA}
\address[CAVS]{Center for Advanced Vehicular Systems, Mississippi State University, Mississippi State, MS 39762, USA}
\address[ARO]{Army Research Laboratory, Weapons and Materials Research Directorate, Aberdeen Proving Ground, MD 21005, United States}

\begin{abstract}

This paper proposes a systematic generalized formulation for calculating both atomic shuffling and shear candidates for a given compound twinning mode in hexagonal closed-packed metals. Although shuffles play an important role in the mobility of twinning dislocations in non-Bravais metallic lattices, their analytical expressions have not been previously derived. The method is illustrated for both flat planes and corrugated planes which are exemplified by \hkl{11-22} and \hkl{10-12}  twinning modes, respectively. The method distinguishes between shuffle displacements and net shuffles. While shuffle displacements correspond to movements between ideal atom positions in the parent and twin lattices, net shuffles comprise contributions from shear on overlying planes which can operate along opposite directions to those of shuffle displacements. Thus, net shuffles in the twinning direction can vanish in a limiting case, as is interestingly the case for those needed in the second plane by the $\mathbf{b_4}$ dislocation candidate in \hkl{11-22} twinning. It is found that while shuffle displacement vectors can be irrational when ${K_1}$ is corrugated, net shuffle vectors are always rational. 

\end{abstract}


\end{frontmatter}

\section{Introduction}

Recent intense consideration of magnesium and titanium by the automotive and aerospace industries as the best metallic lightweight candidates for massive weight reductions have invigorated investigation of the anisotropy and asymmetry of HCP structures. These anisotropy and asymmetry issues are mainly caused by the inability of any close-packed shear deformation mode to provide \hkl<c>-axis deformation. Thus, in the absence of any observed highly glissile pyramidal slip mode, a more difficult non-basal slip direction is required to provide the minimum of five independent slip systems required for arbitrary deformation. The most widely accepted dislocation in the literature possessing a non-zero \hkl<c> component operates on the second-order pyramidal plane \hkl{11-22} with a net Burgers vector equal to $\frac{1}{3}$\hkl<1 1 -2 -3> (\hkl<c+a>)\citep{Bell1957,Price1961a,Price1961b,Rosenbaum1961}. However, the critical resolved shear stress (CRSS) of this dislocation at low temperatures turned out to be higher than that of twinning on any of the pyramidal planes. Hence, various twin modes activate to accommodate either compression or tension of the \hkl<c> axis. Twinning has been correlated with early crack nucleation, and there is a consensus in literature that it reduces ductility of Mg at low temperatures. This limited ductility is currently the main barrier for broader application of Mg in vehicles \citep{Koike2003,Koike2005,Bieler2009,Bieler2009a,Ma2011,Barrett2012}. Because of the diffusional nature of shuffles, some authors ``tacitly'' proposed that appropriate additions of impurities may be effective in hindering shuffles and thus, reducing the propensity to twinning \citep{Nie2003,Williams2002,Wyatt2012}. Furthermore, shuffles have also been found to play a critical role in the mobility of twinning dislocations by narrowing their core width \citep{Serra1988}. Despite the importance of shuffles in twinning, their analytical expressions have not yet been derived.

If the Burgers vector of a compound twinning dislocation in an HCP structure acts on the first plane ($\mathbf{b_1}$) above the $K_1$ composition plane, i.e., the step height is a single interplanar spacing, so shuffles in the twinning direction may not be needed to achieve the mirror symmetry. However, many active twin modes can lower their characteristic shear via growth by disconnections, which have a step height greater than or equal to two interplanar spacings. In fact, the only observed twin mode that ``probably'' has a $\mathbf{b_1}$ type dislocation is \hkl{11-21} extension twinning, which occurs in Ti and Zr. The reduction of the shear magnitude with a dislocation of higher step, however, comes at the expense of increasingly requiring atoms to shuffle along complex paths within the plane of shear \citep{Kiho1958}. Through atomistic simulations, \citet{Serra1988} showed that the core width of active twinning dislocations is noticeably affected by the complexity of atomic shuffles during the passage of the disconnections along the twin boundary.

\citet{Bilby1965} gave an analytical expression for the calculation of shuffles, which required several iterations with guess values to solve. However, the solutions for particular twin modes have never been derived. Moreover, the only available method to derive crystallographically possible Burgers vectors of twinning dislocations is that corresponding to the seminal theoretical analyses by R.C. Pond and co-workers \citep{Pond1979,Pond1985,Serra1991,Pond1994}, which were based on broken translation symmetry and combinations of symmetry operations from each of the two adjacent crystals. \citet{Serra1991a} have used this theory to derive three possible Burgers vectors for \hkl{11-22} twinning. Although comprehensive, this theory has not been adapted yet for shuffles.

In this paper, we fill this current gap by introducing a simple method which allows derivation of shuffles associated with each compound twinning dislocation candidate on any pyramidal plane in HCP metals. This method can also be used to derive the Burgers vectors of these twinning disconnection candidates and corresponding lattice rotations. We illustrate the method for both flat and corrugated pyramidal planes by applying it to \hkl{1-122} and \hkl{10-12} twins in Ti and Mg, respectively.

\section{The general formulation for Shuffles and Burgers vectors}

It should be noted upfront that the current theory was constructed under the hypothesis of compound twinning, which satisfies both Type I and Type II twinning characteristics. It may be valid or extended otherwise, but we have not verified the applicability of the theory for all twins due to the complexity of the problem. However, one must bear in mind that all observed twinning modes in HCP structures that are relevant to current engineering problems are in fact compound twins.  

\begin{figure}[htb!]
\centering
\subfloat{\includegraphics[width = 0.625\textwidth]{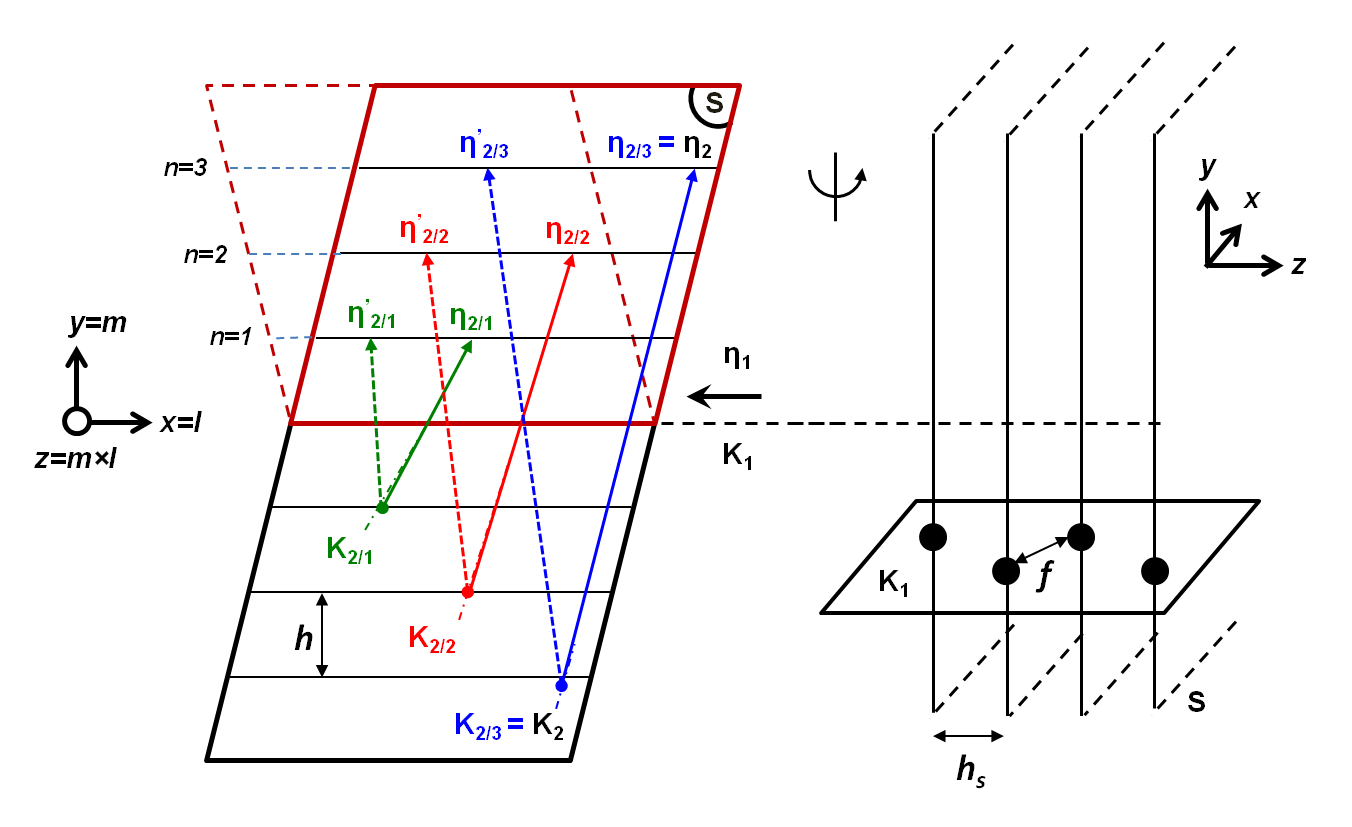}\label{scha}}
\quad
\subfloat{\includegraphics[width = 0.325\textwidth]{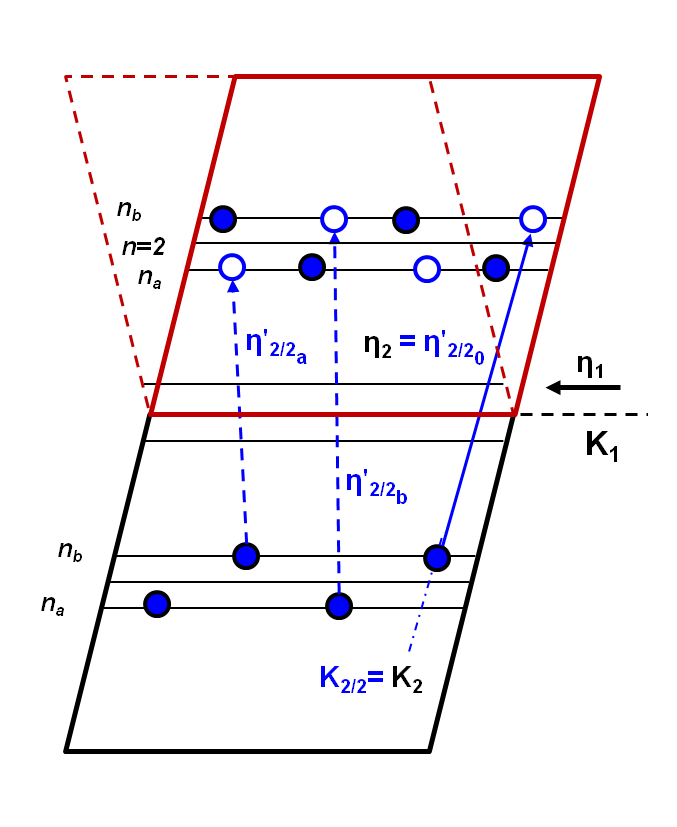}\label{schb}}
\caption{Schematic showing the conventions for shear and shuffle used herein.  The parent and twinned lattices are the solid and dotted parallelograms, respectively, after a shear in the $\mathbf{\eta}_1$ direction on the plane $K_1$.  The flat (left) and corrugated (right) lattices each have multiple possible combinations of $\mathbf{\acute{\eta}}_2$ directions and$K_2$ planes for each $n^{th}$ plane of atoms parallel to the $K_1$ plane (with interplanar spacing of $h$), which are denoted by $\mathbf{\acute{\eta}}_{2/n}$ or $\mathbf{\acute{\eta}}_{2/n_a}$, $\mathbf{\acute{\eta}}_{2/n_b}$, $\mathbf{\acute{\eta}}_{2/n_0}$ for flat or corrugated $K_1$, respectively.  The middle schematic is a view of the shear planes $S$, showing the interplanar spacing between shear planes $h_s$ and the minimum vector joining atoms on neighboring shear planes $f$.  For corrugated structures, the dark and light atoms represent atoms on different shear planes. }
\label{Schematic}
\end{figure}

The notation and variables for the general formulation in the present work are described in the next several paragraphs and are accompanied by the schematic in Figure \ref{Schematic}.  To be consistent with the notation introduced by \citet{Christian1995}, the orthogonal unit vectors in the direction of shear, normal to the composition (twin) plane, and normal to the plane of the shear ($S$) are denoted by $\mathbf{l}=\mathbf{x}=\boldsymbol{\eta_{1}}/\eta_{1}$, $\mathbf{y}=\mathbf{m}$, and $\mathbf{z}=\mathbf{m} \times \mathbf{l}$, respectively.  A given compound twin mode is defined by the composition plane or simply the twinning plane $K_1$ and corresponding shear vector $\boldsymbol{\eta}_1$, a secondary undistorted, but rotated plane $K_2$ and corresponding vector $\boldsymbol{\eta}_2$, and the shear plane $S$ whose normal is a direction common to both $K_1$ and $K_2$ with no rotation ($\mathbf{z}$ direction).  By this definition, notice that the shear plane $S$ also contains both $\boldsymbol{\eta}_1$ and $\boldsymbol{\eta}_2$.  
 
For a given twin mode, several twinning dislocation candidates may exist. Each candidate will operate on a certain $n^{th}$ plane above the final position of the twin boundary and has its own Burgers vector $\mathbf{b}_n$ and accompanying shuffles vectors $\mathbf{d}_p$ ($p \leq n $) which can be systematically and formally expressed.  Formally expressing the shuffle vectors is critical since for twinning dislocations with $n>1$, the shear alone may not be able to exactly bring atoms of planes lying between the twinning dislocation plane and the composition plane to their exact twinned lattice positions (sometimes, shuffles in the $z$ direction are required even for $n=1$ as in the case of \hkl{11-21} twinning \citep{Khater2013}).  Hence, in many cases, shuffles in the $\mathbf{x}$ and/or $\mathbf{z}$ directions may be required in these intermediate planes.  The shuffle in the $\mathbf{x}$ direction on the $n^{th}$ plane may also be a candidate for a Burgers vector, but a twinning dislocation  may not be favored for a few reasons: (1) if the required $\mathbf{b}_n$ is too large for atoms to move in the same sense\footnote{See for example in Section~\ref{compression} the cases of $\mathbf{b}_2$, $\mathbf{b}_5$, and $\mathbf{b}_6$ for \hkl{11-22} twinning} or (2) if the direction of the Burgers vector $\mathbf{b}_n$ opposes the shear direction $\boldsymbol{\eta}_1$.  Also, shuffles in the $\mathbf{x}$ direction might be easier when cooperating with the affine shear of an overlying twinning dislocation.  Therefore, shuffles are often a necessary component in HCP metals if a dislocation candidate is to operate on an overlying $t^{th}$ plane where $t > n$.  Therefore, a criterion must be specified to predict for each $n^{th}$ plane above the $K_1$ plane those planes that can undergo shear from those that cannot (except by shuffles).  A sound theoretical criterion must also simultaneously lead to the magnitude of each corresponding Burgers vector candidate and that of each shuffle vector candidate. Such a theoretical framework has not yet been developed.

For a given twin mode, such a criterion must first recognize that there are multiple crystallographically \emph{admissible} $\boldsymbol{\eta_2}$ vectors for which a shear on the $n^{th}$ plane is able to bring atoms of that plane to the twin positions.  Furthermore, for every $n^{th}$ plane above the final position of the twin boundary, there is an infinite number of mathematically possible Burgers vectors, but our choice herein is restricted to the two smallest Burgers vectors in the shear plane $S$. Without loss of generality, the criterion must be impartial to this detail.   All possible vectors can be considered as trial vectors that shall obey a certain minimization rule, which states and identifies whether or not a possible twinning dislocation on a given $n^{th}$ plane is physically relevant. We denote these trial vectors by $\boldsymbol{\acute{\eta}}_{2/n}$  with varying $n$ for a given twin mode\footnote{The number ``2" in this notation only keeps the classical notation of $\boldsymbol{\eta_2}$}, leaving the unprimed value $\boldsymbol{\eta}_{2/n}$ for the physically possible $\boldsymbol{\eta}_2$ on the $n^{th}$ plane above the final position of the twin boundary. The $\boldsymbol{\eta_2}$ designation itself should be strictly kept for the actually observed one.  For instance, the flat $K_1$ schematic to the left of Figure \ref{Schematic} shows two $\boldsymbol{\acute{\eta}_{2/n}}$ vectors for each $n^{th}$ plane above the $K_1$ plane, but only one is the physically possible $\boldsymbol{{\eta}_{2/n}}$ and an associated $K_{2/n}$ plane (shown as the rightmost vector for simplicity).  However, once the possible $\boldsymbol{{\eta}_{2/n}}$ are identified for each plane, they must be evaluated using a criterion to predict the actual $\boldsymbol{{\eta}_{2}}$ and $K_{2}$ plane (shown as $\boldsymbol{{\eta}_{2/3}}$ and $K_{2/3}$, respectively).  Obviously, in varying the twinning dislocation candidate for a given twin mode, only $\boldsymbol{\eta}_{2}$ and $K_2$ can vary, while $\boldsymbol{\eta_{1}}$ and $K_1$ will remain constant. For each $\boldsymbol{\acute{\eta}}_{2/n}$, the most \emph{possible} dislocation candidate $\mathbf{\acute{b}}_n$ can be envisaged to compete with the most \emph{possible} shuffle displacement  $\mathbf{\acute{d}}_n$. Consequently, the criterion must ascertain whether twinning can be accommodated by $\mathbf{{b}}_n=\mathbf{\acute{b}}_n$, which therefore means that $\boldsymbol{{\eta}}_{2/n}=\boldsymbol{\acute{\eta}}_{2/n}$. Otherwise, this formalized criterion for shear or shuffle on plane $n$ will predict a shuffle $\mathbf{{d}}_n=\mathbf{\acute{d}}_n$.

\subsection{Axiom variables \label{variables}}

Let $\boldsymbol{\acute{\eta}}_{2/n}$ be the smallest lattice vector on the shear plane satisfying the following condition for each $n^{th}$ plane above the final position of the twin boundary:

\begin{equation}
\lVert\boldsymbol{\acute{\eta}}_{2/n} \cdot \mathbf{m}\rVert = 2nh
\label{condition}
\end{equation}

\noindent where $h$ is the average interplanar spacing of the twin plane $K_{1}$.  Equation~\ref{condition} is an obvious requirement for the projection of $\boldsymbol{\eta}_{2/n}$ onto $\mathbf{m}$.  Note that the vector $\boldsymbol{\eta}_{2/n}$ extends from the $-n^{th}$ to the $+n^{th}$ plane (i.e., encompassing $2n$ planes of height $h$).  Meanwhile, to minimize the potential shear on the plane $n$, $\boldsymbol{\eta}_{2/n}$ should achieve the smallest projection onto $\mathbf{x}$. 

Observe that when $K_{1}$ is corrugated, two vectors may satisfy Equation~\ref{condition} on the $n^{th}$ plane if both sub-planes, $n_a$ and $n_b$ located at $nh-(h_{m}/2)$ and $nh+(h_{m}/2)$ are considered, where $h_{m}$ is the smallest interplanar spacing between two different corrugated planes.  That is, $h_m$ is equal to the distance between planes $n_a$ and $n_b$ in Figure \ref{Schematic} ($h_m$ vanishes for a flat $K_{1}$).  In this corrugated case we have:

\begin{equation}
\lVert\boldsymbol{\acute{\eta}}_{2/n_0}\rVert = 2nh, \quad \lVert\boldsymbol{\acute{\eta}}_{2/n_a}\rVert = 2nh - h_{m}, \quad \lVert\boldsymbol{\acute{\eta}}_{2/n_b}\rVert = 2nh + h_{m}
\label{corrugated}
\end{equation}

\noindent where $\boldsymbol{\acute{\eta}}_{2/n_0}$, $\boldsymbol{\acute{\eta}}_{2/n_a}$, and $\boldsymbol{\acute{\eta}}_{2/n_b}$ represent three $\boldsymbol{\acute{\eta}}_{2}$ vectors on the $n^{th}$ plane from (1) the $n_a$ ($n_b$) plane in the parent lattice to the $n_a$ ($n_b$) plane in the twinned lattice, (2) the $n_b$ plane in the parent lattice to the $n_a$ plane in the twinned lattice, and (3) the $n_a$ plane in the parent lattice to the $n_b$ plane in the twinned lattice, respectively.  As can be seen from Equations \ref{condition} and \ref{corrugated}, a distinction must be made between flat and corrugated composition planes to correctly extract the Burgers vector candidates $\mathbf{b}_{n}$ and shuffle displacements $\mathbf{d}_{nx}$, $\mathbf{d}_{ny}$, $\mathbf{d}_{nz}$ for any compound twin mode in an HCP metal.  

The shuffle displacement in the $\mathbf{z}$ direction can be important when considering shuffles.  Let $\mathbf{f}$ be the smallest \emph{lattice} vector in the composition plane joining two neighboring shear planes characterized by an interplanar spacing of $h_{s}$.  The utilization of the vector $\mathbf{f}$  is rather counterintuitive since it is not located within the shear plane, but it is still necessary for considering shuffle along $\mathbf{z}$ which adds to shuffles along the the twinning direction. The definition of the vector $\mathbf{f}$ is dictated by the intriguing characteristic of twinning in HCP metals; all twinning modes without any exception have the prismatic planes, either first or second order, as the shear plane. The theory exposed in next section illustrates how this vector intervenes in the competition between potential shear and potential shuffle on a given plane. 

This distinction is covered in Sections \ref{K1_flat} and \ref{K1_corrugated} for flat and corrugated $K_1$, respectively, and is followed by Section \ref{Net_shuffles}, which details how to calculate the net shuffles required for $n>1$ twinning dislocations.

\subsection{Twinning with $K_1$ flat\label{K1_flat}}

The Burgers vector and shuffle vectors for flat $K_1$ can be systematically calculated given the lattice crystallography.  For twinning in a system with a flat $K_1$, the shuffle in the $\mathbf{y}$ direction, $\mathbf{d}_{ny}$, vanishes since all atoms are located on flat planes parallel to $K_1$.  Therefore, we need to only consider $\mathbf{\acute{b}}_{n}$, $\mathbf{\acute{d}}_{nx}$, and $\mathbf{\acute{d}}_{nz}$, i.e., 

\begin{equation}
\mathbf{\acute{b}}_{n}= \sgn\left(\boldsymbol{\acute{\eta}}_{2/n} \cdot \frac{\boldsymbol{\eta}_{1}} {\eta_{1}}\right) \left(\sqrt{\acute{\eta}^{2}_{2/n}-4n^{2}h^{2}}\right)\frac{\boldsymbol{\eta}_{1}} {\eta_{1}}
\label{shearf}
\end{equation}

\begin{equation}
\mathbf{\acute{d}}_{nx}= \sgn\left(\boldsymbol{\acute{\eta}}_{2/n} \cdot \frac{\boldsymbol{\eta}_{1}} {\eta_{1}}\right) \left(\sqrt{\acute{\eta}^{2}_{2/n}-4n^{2}h^{2}} - \Bigl|\mathbf{f} \cdot \frac{\boldsymbol{\eta}_{1}} {\eta_{1}}\Bigr| \right)\frac{\boldsymbol{\eta}_{1}} {\eta_{1}}
\label{shufflefx}
\end{equation}

\begin{equation}
\mathbf{\acute{d}}_{nz}= \pm h_{s} \mathbf{z}
\label{shufflefz}
\end{equation}

\noindent where $\eta_1$ and $\eta_{2/n}$ are the magnitudes of vectors $\boldsymbol{\eta}_{1}$ and $\boldsymbol{\eta}_{2/n}$, respectively.  In Equation \ref{shearf}, the first term describes whether the twinning dislocation will act in the $+\boldsymbol{\eta}_{1}$ or $-\boldsymbol{\eta}_{1}$ direction, the second term (in parenthesis) describes the Burgers vector magnitude of the twinning dislocation, and the last term is a unit vector in the $\boldsymbol{\eta}_{1}$ direction.  Equations \ref{shufflefx} and \ref{shufflefz} describe the orthogonal shuffle vectors in the $\boldsymbol{\eta}_{1}$ and $\mathbf{z}$ directions, whereby (i) the magnitude along $\boldsymbol{\eta}_{1}/\eta_1$ is obtained from the projection of $\mathbf{f}$ onto the $\boldsymbol{\eta}_{1}$ unit vector and the previously calculated Burgers vector magnitude (Eq.~\ref{shearf}) and (ii) the magnitude along $\mathbf{z}$ is merely related to the interplanar spacing $h_s$ of two neighboring shear planes.  For $\mathbf{\acute{d}}_{nx}$, when the $\mathbf{f} \cdot \boldsymbol{\eta}_{1}$ term is greater than the twinning dislocation magnitude (as in the flat $K_1$ case), the shuffle  $\mathbf{\acute{d}}_{nx}$ will always be in the opposite direction of the shear  $\mathbf{\acute{b}}_n$.  When the $\mathbf{f} \cdot \boldsymbol{\eta}_{1}$ term is equal to zero (as in the corrugated $K_1$ case), the shuffle $\mathbf{\acute{d}}_{nx}$ is equal to the shear $\mathbf{\acute{b}}_n$ in magnitude and directionality.

The criterion used to identify whether a twinning dislocation or whether shuffles act on plane $n$ is as follows.  If the Burgers vector magnitude of the twinning dislocation is less than the shuffle vector magnitude in the $\mathbf{x}$ direction (i.e., $\lVert\mathbf{\acute{b}}_n\rVert \leq \lVert\mathbf{\acute{d}}_{nx}\rVert$), then $\mathbf{b}_n=\mathbf{\acute{b}}_n$ is a twinning dislocation candidate operating on the $n^{th}$ plane, and $\boldsymbol{\acute{\eta}}_{2/n}$ is the actual $\boldsymbol{\eta}_{2/n}$. However, in the opposite case whereby the shuffle vector magnitude is less than the twinning dislocation magnitude (i.e., $\lVert\mathbf{\acute{b}_n}\rVert > \lVert\mathbf{\acute{d}}_{nx}\rVert$), then no twinning dislocation can operate on the $n^{th}$ plane.  Instead, shuffle displacements $\mathbf{d}_{nx}=\mathbf{\acute{d}}_{nx}$ and $\mathbf{d}_{nz}=\mathbf{\acute{d}}_{nz}$ must take place in the $\mathbf{x}$ and $\mathbf{z}$ directions, respectively, for any twinning dislocation $\mathbf{b}_t$ which is operational on the $t^{th}$ plane such that $t>n$.  Notice that $t$ is used to differentiate the plane of the twinning dislocation from the underlying planes $n<t$.  Restated in a slightly different way, a twinning dislocation $\mathbf{b}_t$ on plane $t$ may require shuffles $\mathbf{d}_{n}$ on the underlying planes $n=\left\{1,~\ldots{,~t-1}\right\}$, but the influence of $\mathbf{b}_t$ on these planes must also be taken into account (i.e., the net shuffles in Section \ref{Net_shuffles}).  It should also be mentioned that this criterion initially only compares the shear magnitude with the shuffle magnitude in $\mathbf{x}$; the shuffles in the $\mathbf{z}$ direction was deemed less important, but may also play a role in this predictive criterion.

\subsection{Twinning with $K_1$ corrugated\label{K1_corrugated}}

In a similar manner to Section \ref{K1_flat}, the Burgers vector and shuffle vectors for a corrugated $K_1$ can also be calculated given the lattice crystallography.  The dot product $\left(\mathbf{f} \cdot {\boldsymbol{\eta}_{1}}/{\eta_{1}}\right)$ vanishes for all twin modes observed in HCP structures having $K_1$ corrugated. In fact, the shear plane for all these modes belong to the \hkl{-12-10} family. This interesting characteristic implies that the vector $\mathbf{f}$ is not needed for shuffle derivation. Without loss of generality, one may still use Equations~\ref{shearf} and \ref{shufflefx} from the flat $K_1$ plane to calculate the twinning dislocation Burgers vector and the shuffle in the $\mathbf{x}$ direction, but Equation \ref{shufflefz} for $\mathbf{d}_{nz}$ is equal to zero. 

Recall that there are three potential $K_2$ planes and $\boldsymbol{\acute{\eta}}_{2/n}$ vectors for the $n^{th}$ plane in the corrugated $K_1$ twinning mode: $\boldsymbol{\acute{\eta}}_{2/n_0}$, $\boldsymbol{\acute{\eta}}_{2/n_a}$, and $\boldsymbol{\acute{\eta}}_{2/n_b}$.  Hence, consider the vectors $\mathbf{\acute{b}}_{n_0}$, $\mathbf{\acute{b}}_{n_a}$, and $\mathbf{\acute{b}}_{n_b}$ for each of $\boldsymbol{\acute{\eta}}_{2/n_0}$, $\boldsymbol{\acute{\eta}}_{2/n_a}$, and $\boldsymbol{\acute{\eta}}_{n_b}$, respectively, to be:

\begin{equation}
\mathbf{\acute{b}}_{n_0}= \sgn\left(\boldsymbol{\acute{\eta}}_{2/n} \cdot \frac{\boldsymbol{\eta}_{1}} {\eta_{1}}\right)\left(\sqrt{\acute{\eta}^{2}_{2/n}-4n^{2}h^{2}}\right)\frac{\boldsymbol{\eta}_{1}} {\eta_{1}}
\label{shearc0}
\end{equation}

\begin{equation}
\mathbf{\acute{b}}_{n_a}= \sgn\left(\boldsymbol{\acute{\eta}}_{2/n_a} \cdot \frac{\boldsymbol{\eta}_{1}} {\eta_{1}}\right) \left(\sqrt{\acute{\eta}^{2}_{2/n_a}-\left(2nh-h_{m}\right)^{2}}\right)\frac{\boldsymbol{\eta}_{1}} {\eta_{1}}
\label{shearca}
\end{equation}

\begin{equation}
\mathbf{\acute{b}}_{n_b}= \sgn\left(\boldsymbol{\acute{\eta}}_{2/n_b} \cdot \frac{\boldsymbol{\eta}_{1}} {\eta_{1}}\right) \left(\sqrt{\acute{\eta}^{2}_{2/n_b}-\left(2nh+h_{m}\right)^{2}}\right)\frac{\boldsymbol{\eta}_{1}} {\eta_{1}}
\label{shearcb}
\end{equation}

\begin{equation}
\mathbf{\acute{d}}_{ny}= \pm h_{m} \mathbf{y}
\label{shufflecy}
\end{equation}

A criterion is again used to identify whether a twinning dislocation or whether shuffles act on plane $n$.  If the Burgers vector magnitude of the twinning dislocation for $\boldsymbol{\acute{\eta}}_{2/n_0}$ is less than the Burgers vector magnitude for $\boldsymbol{\acute{\eta}}_{2/n_a}$ (i.e., $\lVert\mathbf{\acute{b}}_{n_0}\rVert < \lVert\mathbf{\acute{b}}_{n_a}\rVert$), then $\mathbf{b}_n = \mathbf{\acute{b}}_{n_0}$ is a twinning dislocation candidate that shifts atoms in both the $n_a$ and $n_b$ planes in the twinned lattice to mirror atom positions in the parent lattice with respect to the $\mathbf{x}$ direction.  However, the atom positions in terms of the $\mathbf{y}$ direction are not reversed, so this twinning dislocation must be accompanied by a shuffle $\mathbf{d}_{ny}=\mathbf{\acute{d}}_{ny}= \pm h_{m} \mathbf{y}$ that will shuffle atoms  in the twinned lattice from the $n_a$ plane to the $n_b$ plane, and vice versa, to mirror the parent lattice about the $K_1$ composition plane. Furthermore, shuffles with alternating sign $\mathbf{d}_{ny}=\mathbf{\acute{d}}_{ny}$ must operate within the $K_1$ composition plane. On the other hand, if the Burgers vector magnitude of the twinning dislocation for $\boldsymbol{\acute{\eta}}_{2/n_0}$ is greater than or equal to the Burgers vector magnitude for $\boldsymbol{\acute{\eta}}_{2/n_a}$ (i.e., $\lVert\mathbf{\acute{b}}_{n_0}\rVert \geq \lVert\mathbf{\acute{b}}_{n_a}\rVert$), no twinning dislocation could operate on this plane, and instead two different shuffles in the $\mathbf{x}$ direction must operate on the sub-planes $n_a$ and $n_b$ for any twinning dislocation $\mathbf{b}_t$ such that $t>n$. These shuffles are $\mathbf{d}_{n_ax}=\mathbf{\acute{b}}_{n_a}$ and $\mathbf{d}_{n_bx}=\mathbf{\acute{b}}_{n_b}$.

\subsection{Net shuffles\label{Net_shuffles}}

In the preceding section, we defined shuffle displacements $\mathbf{d}_{nx}$ on the $n^{th}$ plane in the $\mathbf{x}$ direction as the displacements necessary for an atom to move from its ideal parent lattice position to its ideal twin lattice position excluding any contribution of shear. However, this is independent of the shear from the twinning dislocation, which needs to be accounted for in the case of disconnections.  For instance, if a twinning dislocation acts on the $4^{th}$ plane to bring these lattice atoms into their twinned lattice positions, it will also shear atoms on the $1^{st}$, $2^{nd}$, and $3^{rd}$ planes.  Hence, the atoms on these underlying planes may require shuffles to bring them to their twinned lattice positions.  The difference between the shuffle vectors required for these atoms on these underlying planes and the shear from the twinning dislocation is the net shuffle that is required.  The first step in calculating the net shuffle is to calculate the shear on these planes.  The shear from any twinning dislocation $\mathbf{b}_t$ on the $t^{th}$ plane will motivate atoms in the underlying planes ($n<t$) to move by an amount equal to:

\begin{equation}
\mathbf{d}_s=\frac{nh \pm (h_m/2)}{th}\mathbf{b}_t
\end{equation} 

\noindent where $n$ is for planes less than $t$.  In the case of flat $K_1$, this equation reduces to $\mathbf{d}_s=\left({n}/{t}\right)\mathbf{b}_t$.  Thus, the twinning dislocation shear may decrease or increase the shuffles needed to produce the twin lattice. So, in a limiting case, it is perfectly possible that no shuffles would be needed if the shear caused by a $\mathbf{b}_t$ dislocation is able to bring all atoms of an underlying plane to their correct twin positions (e.g., the shuffles on the $2^{nd}$ plane are exactly equal to $\mathbf{d}_s$ on this plane for the $\mathbf{b}_4$ dislocation candidate in \hkl{11-22} twinning). Therefore, as shuffles correspond to completely local rearrangements of atoms which do not cause plastic strain, the homogeneous shear of the crystal should be subtracted from the shuffle displacements to now define the true or net shuffles $\mathbf{\Delta}^{t}_{n}$:

\begin{equation}
\mathbf{\Delta}^{t}_{n}=\mathbf{d}_{n}-\mathbf{d}_s =\mathbf{d}_{n} - \frac{nh \pm (h_m/2)}{th}\mathbf{b}_t
\label{netshuffle}
\end{equation}

The derivation of net shuffles $\mathbf{\Delta}^{t}_{n}$ is necessary to distinguish between the work of deviatoric stress and that of hydrostatic pressure in the core of the twinning dislocation \citep{Ostapovets2013}. Since net shuffles correspond to pure diffusion of atoms, net shuffles are not driven by any deviatoric stress and should rather be caused by hydrostatic pressure, as diffusion is well-known to depend on hydrostatic pressure.  This observation is of paramount importance because it elucidates why twin nucleation was experimentally \citep{Christian1995} and numerically \citep{Barrett2012} observed to depend on stress concentration, i.e., triaxiality. This definition has a further advantage of being able to elucidate symmetries in the twinning mechanisms not otherwise possible.  

\subsection{Mathematical scheme for orthonormal four-index vectors \label{math}}

The classical non-orthonormal Miller-Bravais notations is used herein to identify directions and planes.  However, a limitation of using the non-orthonormal Miller-Bravais vectors that has discouraged its use is that vector operations are not as straightforward as with the more conventional orthonormal three-index vector space.  Clearly, the Miller-Bravais system more easily displays the symmetry of the HCP lattice, though.  A powerful method which benefits from both four-index notations and an orthonormal basis was proposed and used by \citet{Pond1987a,Pond1987b,Pond1991,Serra1986,Serra1988,Serra1991} as a modification of Frank's powerful four-vector extension of Miller-Bravais by projection from a higher dimension.  Herein, we utilize the same four-vector space to calculate shear and shuffles.

First, to appropriately transform the non-orthonormal Miller-Bravais vector $\left[uvtw\right]$ into the orthonormal four-space vector $\left[uvt\Lambda{w}\right]_4$ requires the use of a factor $\Lambda=\left(2/3\right)^{1/2}\gamma$ where $\gamma=c/a$ ratio.  Additionally, the four-space vectors have a magnitude of $e=\left(3/2\right)^{1/2}a$ where $a$ is the lattice parameter.  Utilizing this transformation allows crystallographic calculations and vector operations to be carried out in a rather straightforward manner \cite{Pond1987a,Pond1987b,Pond1991,Serra1986,Serra1988,Serra1991}.  For example, the vector magnitude and vector scalar product is given by

\begin{equation}
\lVert\left[uvt{w}\right]\rVert=\lVert\left[uvt\Lambda{w}\right]_4\rVert=\left({u^2 + v^2 + t^2 + \Lambda^2w^2}\right)^{1/2}e
\label{vector1}
\end{equation}

\noindent and 

\begin{equation}
\left[u_1v_1t_1{w}_1\right]\cdot\left[u_2v_2t_2w_2\right]=\left[u_1v_1t_1\Lambda{w}_1\right]_4 \cdot \left[u_2v_2t_2\Lambda{w}_2\right]_4 = \left( {u_1u_2 + v_1v_2 + t_1t_2 + \Lambda^2w_1w_2} \right) e^2
\label{vector2}
\end{equation}

\noindent Additionally, from these expressions, the angle can be easily calculated using the traditional form, $\mathbf{a}\cdot\mathbf{b}=\lVert\mathbf{a}\rVert \lVert\mathbf{b} \rVert \textrm{cos } \theta$.  The other useful vector operation is to calculate the interplanar spacing for a plane $\left(hkil\right)$, i.e., 

\begin{equation}
d = \left({{h^2 + i^2 + k^2 + l^2/\Lambda^2}}\right)^{-1/2}e
\label{vector3}
\end{equation}

Now that both the formalized theory and the mathematics have been laid out, this can be applied to HCP structures.  In the following, the above theory which formally describes the Burgers vector candidates and their corresponding shuffles is applied to \hkl{11-22} and \hkl{10-12} twin modes in HCP structures.

\section{Candidacy in \hkl{11-22} twinning \label{compression}}

\begin{figure}[ht]
\centering
\includegraphics[width = \textwidth]{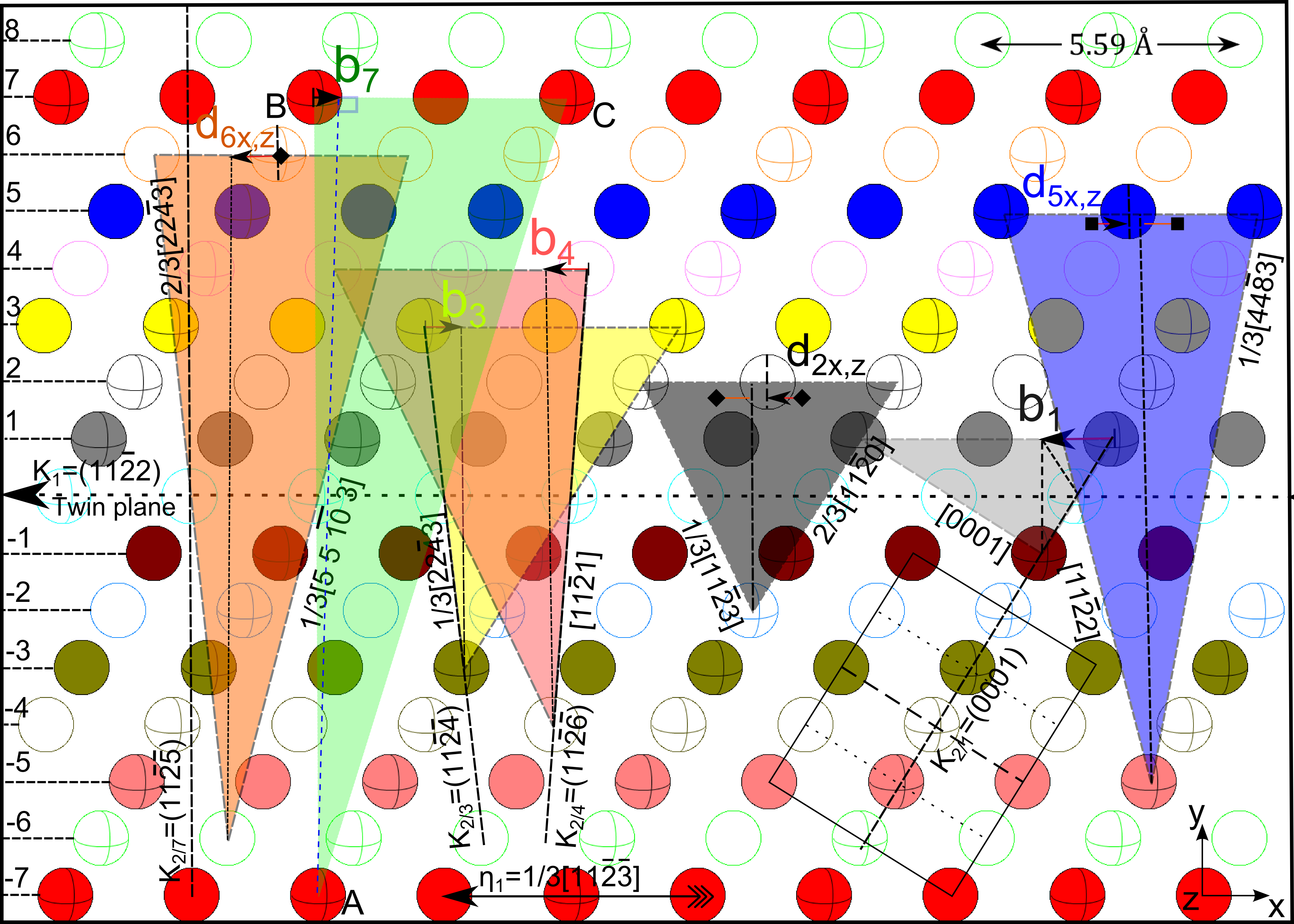}
\caption{\hkl(-1 1 0 0) shear-plane view of atom positions in pure Ti constructed by CrystalMaker\textsuperscript{\textregistered} \citep{CrystalMaker} displaying 16 layers of \hkl(1 0 -1 2) planes which reveal the crystallography of shear and shuffles of \hkl(1 1 -2 2)\hkl[1 1 -2 -3] twinning. Without loss of generality, the final position of the twin boundary for each twinning dislocation candidate is assumed to be at the $0^{th}$ plane which is the eighth plane counting from the bottom plane. The atoms above each plane of a given operation $\mathbf{b}_n$ correspond to the untwinned parent lattice and thus should be disregarded when one ponders about the effect of each $\mathbf{b}_n$.}
\label{Pmaster}
\end{figure}

\begin{figure}[ht]
\centering
\includegraphics[width = 0.75\textwidth]{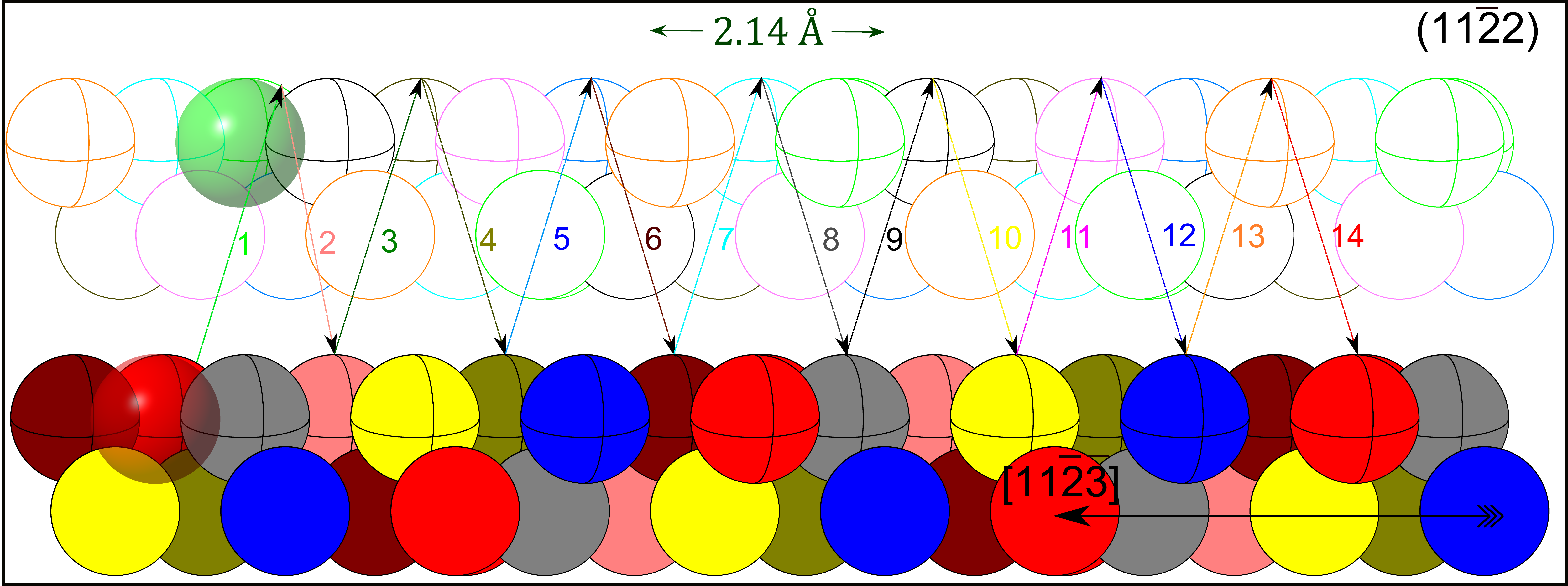}
\caption{A \hkl(1 1 -2 2) view of atom positions in pure Ti constructed by CrystalMaker\textsuperscript{\textregistered} \citep{CrystalMaker} displaying four layers of \hkl{-1 1 0 0} planes revealing the stacking sequence of \hkl{1 1 -2 2} planes in HCP structures.}
\label{Prmaster}
\end{figure}

For these flat planes, the seventh $\boldsymbol{\acute{\eta}}_{2/7}$ leads to an extremely small possible Burgers vector, $\mathbf{b}_7$. This deviation is better viewed in the stacking sequence of the shear plane reported in Figure~\ref{Prmaster}. As such, it is sufficient to consider $\boldsymbol{\acute{\eta}}_{2/n}$ only up to $n=7$. The small size of this vector did allow proper illustration of $\mathbf{b}_7$ in Figure~\ref{Pmaster} as the corresponding $\boldsymbol{\acute{\eta}}_{2/n}$ had to be exceedingly inclined.  Thus, to elucidate the size of this vector we reported in Figure~\ref{twinview} a top view of planes $n=-7$ and $n=7$, where it can be clearly seen that \hkl{11-22} planes show an extremely small deviation from an exact stacking sequence on the $14^{th}$ plane.  A number of observations should be pointed out in Figure~\ref{Pmaster} to help supplement the formalized theory.

\begin{enumerate}

	\item \textbf{Lattice and twin boundary}. To illustrate the theory, only atom positions of the parent are included. In fact, a sound theory must be capable of predicting the required movements starting from an untwinned arrangement of atoms. The position of $K_1$ ($0^{th}$ plane) corresponds to the position of the twin boundary after the activation of a given $\mathbf{b}_n$ twinning dislocation.  All atom positions above the plane on which the $\mathbf{b}_n$ twinning dislocation passes correspond to the parent lattice and must be disregarded when predicting a given $\mathbf{b}_n$.  
	\item \textbf{Atom color}.  The atoms are colored to reflect the stacking of \hkl{11-22} planes in the $\mathbf{y}$ direction.  These planes alternate between filled and unfilled as a guide to the eye.  For example, in Figure~\ref{Pmaster} and later in the corrugated case, atoms of the same color belong to the same plane, which is parallel to $K_1$. 
	\item  \textbf{Atom style}.  The atom style (filled versus unfilled, plain versus equatorial) reflects the $ABCD$ stacking sequence of the \hkl(-1100) shear plane (into the page). The four styles are hollow, shaded, equatorial hollow, and equatorial shaded circles.  Examine the box drawn in the lower right hand corner of Figure~\ref{Pmaster}.  The equatorial and simple circles reflect the the $AB$ stacking sequence of the basal plane. The shaded and hollow atoms each belong to a sequence of two closely stacked \hkl{-1100} plane pairs. This is illustrated in Figure~\ref{Prmaster} which is a \hkl(11-22) projection of the \hkl{-1100} planes. 
	\item \textbf{Plane numbering convention}.   The plane count starts from an atom (in red, $-7$ plane) which is situated at the lowest \hkl(11-22) plane in Figure~\ref{Pmaster} and then proceeds to the next plane above. The numbers on the left hand side in Figure~\ref{Prmaster} highlight this count. At plane $\#$14 (in red, $+7$) the position of the first atom is nearly reproduced by the red atom at plane $\#$1 (in red, $-7$) drawn with the same style.  The transition from $-n$ planes to $+n$ planes occurs at the $K_1$ composition plane, which is the final position of the twin boundary should any highlighted dislocation candidate be operational.  Thus, the numbering convention chosen here is used to highlight that atoms in $+n$ plane need to shear and/or shuffle to mirror the atom positions in the $-n$ plane.
	\item \textbf{Triangles}.  The triangles are drawn to show the most likely shear and shuffle candidates for each plane.  First, an atom from each $-n$ plane is selected.  Then, a line perpendicular to the $K_1$ plane is drawn to show the necessary position of a mirror atom in the twinned lattice on the $+n$ plane.  Now, the corner atoms of the triangle on the $+n$ plane are the closest atoms in the $-\boldsymbol{{\eta}}_n$ and $+\boldsymbol{{\eta}}_n$ directions from the mirror atom that are on the same shear plane as the initial $-n$ atom (i.e., same style: hollow, shaded, equitorial, etc.).
	\item $K_2$ \textbf{and} $\boldsymbol{{\eta}}_n$.  Many of the $\boldsymbol{{\eta}}_n$ are shown along the sides of the shaded triangles (in some cases, only the relevant $\boldsymbol{{\eta}}_n$ is shown).  Additionally, the $K_2$ planes are shown for the four twinning dislocation candidates on the $1^{st}$, $3^{rd}$, $4^{th}$, and $7^{th}$ planes.
	\item \textbf{Shear and shuffle vectors}.  The shear $\mathbf{b}_n$ and shuffle $\mathbf{d}_n$ vectors are shown for the different planes.  To understand whether it is more likely for shear or shuffle to operate, examine the three atoms along the $+n$ plane.  First, the twinning dislocation candidate can be evaluated.  The distance from the corner atoms of the triangle to the twinned atom position is the Burgers vector magnitude of the twinning dislocation; the smaller the Burgers vector, the more likely it is that the twinning dislocation is the operative one for that plane.  There is also a directionality based on this Burgers vector.  Notice that $\mathbf{b}_1$ and $\mathbf{b}_4$ are in an opposite direction of $\mathbf{b}_3$ and $\mathbf{b}_7$; the former directions are more likely in $c$-axis tension and the latter directions in $c$-axis compression.  Second, the shuffle vector can be evaluated from the triangle.  In this case, the distance from the middle atom on the $+n$ plane to the atom position in the twinned lattice is calculated.  Notice that this atom is on a different shear plane (i.e., different atom style) than the atom lying on the triangle vertice on the $-n$ plane.  Hence, a shuffle is required in both the $\mathbf{x}$ and $\mathbf{z}$ directions.  The minimum Burgers vector or shuffle vector magnitude in the $\mathbf{x}$ direction is the operable twinning dislocation or shuffle in that plane, respectively.  This is denoted by the presence of $\mathbf{b}_n$ or $\mathbf{d}_n$ above the shaded triangle in Figure~\ref{Pmaster}. When there is a diamond or a square symbol on a given $\mathbf{d}_n$ vector, additional movements are required along the $z$-axis in the positive and negative senses, respectively. Realize that, in some cases, the twinning dislocation may actually act as a shuffle for a twinning dislocation on an overlying plane.
\end{enumerate}

Therefore, Figure~\ref{Pmaster} pictorially captures much of the formalized theory for calculating shears and shuffles for flat $K_1$ in HCP metals.  For calculating the analytical expressions, though, the height of \hkl{11-22} planes is equal to $ h=a \gamma/(2\sqrt{\gamma^2 + 1})$ and the seven $\boldsymbol{\acute{\eta}}_{2/n}$ vectors satisfying Equation~\ref{condition} for $n=1,...,7$ are found to be $1/3$\hkl[-1 1 0 0], $1/3$\hkl[-1 -1 2 -3], $1/3$\hkl[-2 -2 4 -3], \hkl[-1 -1 2 -1], $1/3$\hkl[-4 -4 8 -3], $2/3$\hkl[-2 -2 4 -3], and $1/3$\hkl[-5 -5 10 -6], respectively. For each of these vectors, a section was dedicated where the Burgers vectors and shuffles are computed through the systematic use of Equations~\ref{shearf}-\ref{shufflefz}. The vector $\mathbf{f}$ for the \hkl{11-22} planes could be either $1/6$\hkl[2 0 -2 -3] or $1/6$\hkl[0 2 -2 -3] (same result), which is illustrated in Figure~\ref{ztri}.  The $\boldsymbol{\eta}_{1}$ direction is $1/3$\hkl[1 1 -2 -3].

\subsection{Case of $\boldsymbol{\acute{\eta}}_{2/1}=1/3\hkl[-1 -1 2 0]$ }

Substituting $\boldsymbol{\acute{\eta}}_{2/1}=1/3\hkl[-1 -1 2 0]$ and $\mathbf{f}=1/6\hkl[2 0 -2 -3]$ in Equations~\ref{shearf}-\ref{shufflefz}, we obtain $\lVert\mathbf{\acute{b}_1}\rVert = \lVert\mathbf{\acute{d}_{1x}}\rVert$, and as such, $\mathbf{b}_1$ with $\boldsymbol{\eta}_{2/1}=1/3\hkl[-1 -1 2 0]$, and $K_{2/1}=(0 0 0 1)$, is a possible twinning dislocation candidate accommodating tension of the \hkl<c>-axis. The Burgers vector is given by Equation~\ref{shearf} to be:

\begin{equation}
\mathbf{b}_{1} = \acute{\mathbf{b}}_1= \frac{1}{3(\gamma^{2}+1)}[11\bar{2}\bar{3}]= \frac{1}{\gamma^{2}+1}\boldsymbol{\eta}_{1},\quad 
{b_{1}} = \frac{a}{\sqrt{\gamma^{2}+1}}
\end{equation}

\noindent   The geometrical implications of the theory reduces to classical trigonometry applied to the light gray triangle highlighted in Figure~\ref{Pmaster}.  In this equation,  $\boldsymbol{\eta}_{1}$ is included to show that the twinning dislocation direction is in the same direction as $\boldsymbol{\eta}_{1}$.  In the subsequent subsections, notice that the twinning dislocations and shuffles are vectors in either the $+\boldsymbol{\eta}_{1}$ or $-\boldsymbol{\eta}_{1}$ directions (in $\mathbf{x}$).  Additionally, the shear and shuffle directions in $\mathbf{x}$ are always opposite, as can be clearly seen from the triangles in Figure~\ref{Pmaster}.

\subsection{Case of $\boldsymbol{\acute{\eta}}_{2/2}=1/3\hkl[-1 -1 2 -3]$}
\label{d2}

Substituting $\boldsymbol{\acute{\eta}}_{2/2}=1/3\hkl[-1 -1 2 -3]$ and $\mathbf{f}=1/6\hkl[2 0 -2 -3]$ in Equations~\ref{shearf}-\ref{shufflefz}, we obtain: 

\begin{equation}
\acute{\mathbf{b}}_{2} = \frac{\gamma^{2}-1}{3(\gamma^{2}+1)}[\bar{1}\bar{1}23]
\end{equation}

\begin{equation}
\acute{\mathbf{d}}_{2x} = \frac{(3-\gamma^{2})}{6(\gamma^{2}+1)}[11\bar{2}\bar{3}] 
\end{equation}

Therefore $\lVert\mathbf{\acute{b}}_2\rVert > \lVert\mathbf{\acute{d}}_{2x}\rVert$, and as such, no twinning dislocation can glide on this plane which will be instead acted on by shuffles in the $\mathbf{x}$ and $\mathbf{z}$ directions with the following displacements for any overlying dislocation:

\begin{equation}
\mathbf{d}_{2x} = \acute{\mathbf{d}}_{2x}, \quad d_{2x} = \frac{a}{2}\frac{|3-\gamma^{2}|}{\sqrt{\gamma^{2}+1}}
\end{equation}

\begin{equation}
\mathbf{d}_{2z} = \mp\frac{1}{6}[\bar{1}100], \quad d_{2z}= \frac{a\sqrt{3}}{6}
\end{equation}

\begin{figure}[ht]
\centering
\includegraphics[width = 0.75\textwidth]{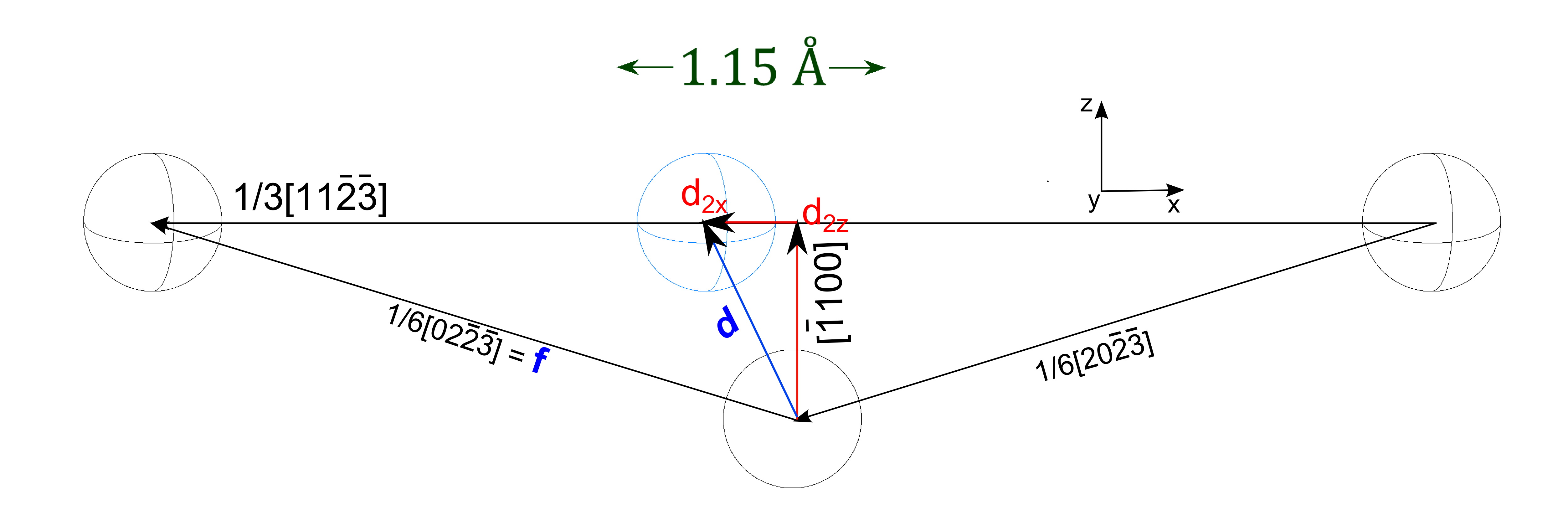}\label{Shufflez}
\caption{\label{ztri} A \hkl{1 1 -2 2} view illustrating shuffle displacements undergone by atoms at the second plane along $x$ and $z$ directions, which correspond to 1/3\hkl[1 1 -2 -3] and \hkl[-1 1 0 0], respectively. The twinned position is shown as the blue equitorial atom and the shuffle vector $\mathbf{d}$ is highlighted in blue.}
\end{figure}

The geometrical implications of the theory reduce to trigonometric calculations carried out on both the dark gray triangle in Figure~\ref{Pmaster} and the triangle in Figure~\ref{ztri}.  The shuffle in the $\mathbf{z}$ direction stems from Pythagorean's theorem ($a^2+b^2=c^2$) utilizing the $1/6$\hkl[0 2 -2 -3] and $1/3\hkl[1 1 -2 -3]$ directions to calculate the magnitude of $\mathbf{d}_{2z}$ in the \hkl[-1 1 0 0] direction, i.e., $a=1/6\hkl[1 1 -2 -3]$, and $c=1/6\lVert\hkl[2 0 -2 -3]\rVert$.

\subsection{Case of $\boldsymbol{\acute{\eta}}_{2/3}=1/3\hkl[-2 -2 4 -3]$ }

Substituting $\boldsymbol{\acute{\eta}}_{2/3}=1/3\hkl[-2 -2 4 -3]$ and  $\mathbf{f}=1/6\hkl[2 0 -2 -3]$ in Equations~\ref{shearf}-\ref{shufflefz}, we obtain:

\begin{equation}
\acute{\mathbf{b}}_{3} = \frac{\gamma^{2}-2}{3(\gamma^{2}+1)}[\bar{1}\bar{1}23]
\end{equation}

\begin{equation}
\acute{\mathbf{d}}_{3x} = \frac{5-\gamma^{2}}{6(\gamma^{2}+1)}[11\bar{2}\bar{3}]
\end{equation}

Therefore, $\lVert\mathbf{\acute{b}}_3\rVert < \lVert\mathbf{\acute{d}}_{3x}\rVert$, and as such, $\mathbf{b}_3$ with $\boldsymbol{\eta}_{2/3}$=1/3\hkl[-2 -2 4 -3], and $K_{2/3}=\hkl(1 1 -2 -4)$, is a possible twinning dislocation candidate accommodating compression of the \hkl<c>-axis (as all HCP metals show $\gamma >\sqrt{2}$). The Burgers vector is given from above to be:

\begin{equation}
\mathbf{b}_{3} = \acute{\mathbf{b}}_3, \quad b_{3}= a\frac{|\gamma^{2}-2|}{\sqrt{\gamma^{2}+1}}
\end{equation}

The geometrical implications of the theory are illustrated by the yellow triangle in Figure~\ref{Pmaster}.

\subsection{Case of $\boldsymbol{\acute{\eta}}_{2/4}=\hkl[-1 -1 2 -1]$}

Substituting $\boldsymbol{\acute{\eta}}_{2/4}=\hkl[-1 -1 2 -1]$ and $\mathbf{f}=1/6\hkl[2 0 -2 -3]$ in Equations~\ref{shearf}-\ref{shufflefz}, we obtain:

\begin{equation}
\acute{\mathbf{b}}_{4} = \frac{3-\gamma^{2}}{3(\gamma^{2}+1)}[11\bar{2}\bar{3}] = 2\mathbf{d}_{2x} 
\end{equation}

\begin{equation}
\acute{\mathbf{d}}_{4x} = \frac{3\gamma^{2}-5}{6(\gamma^{2}+1)}[\bar{1}\bar{1}23]
\end{equation}

Therefore, $\lVert\mathbf{\acute{b}}_4\rVert < \lVert\mathbf{\acute{d}}_{4x}\rVert$, and as such, $\mathbf{b}_4$ with $\boldsymbol{\eta}_{2/4}$=\hkl[-1 -1 2 -1], and $K_{2/4}=\hkl(1 1 -2 -6)$, is a possible twinning dislocation candidate accommodating tension/compression of the \hkl<c>-axis for $\gamma$ smaller/greater than $\sqrt{3}$, respectively. The Burgers vector is given from above by

\begin{equation}
\mathbf{b}_{4} = \acute{\mathbf{b}}_{4},\quad b_{4} = a\frac{|3-\gamma^{2}|}{\sqrt{\gamma^{2}+1}}
\end{equation}

The geometrical implications of the theory simplify to basic trigonometry on the pink triangle in Figure~\ref{Pmaster}.  Interestingly, the $\mathbf{b}_4$ twinning dislocation is exactly twice the magnitude of the shear in $\mathbf{x}$ on the 2$^{nd}$ plane (i.e., $\mathbf{d}_{2x})$, which results in a net shuffle of 0 on this plane.

\subsection{Case of $\boldsymbol{\acute{\eta}}_{2/5}=1/3\hkl[-4 -4 8 -3]$}
\label{d5}

Substituting $\boldsymbol{\acute{\eta}}_{2/5}=1/3\hkl[-4 -4 8 -3]$ and $\mathbf{f}=1/6\hkl[2 0 -2 -3]$ in Equations~\ref{shearf}-\ref{shufflefz}, we obtain 

\begin{equation}
\acute{\mathbf{b}}_{5} = \frac{4-\gamma^{2}}{3(\gamma^{2}+1)}[11\bar{2}\bar{3}]
\end{equation}

\begin{equation}
\acute{\mathbf{d}}_{5x} = \frac{3\gamma^{2}-7}{6(\gamma^{2}+1)}[\bar{1}\bar{1}23]
\end{equation}

Therefore, $\lVert\mathbf{\acute{b}}_5\rVert > \lVert\mathbf{\acute{d}}_{5x}\rVert$, and as such, no twinning dislocation can glide on this plane which will instead be acted on by shuffles in the $\mathbf{x}$, and $\mathbf{z}$ directions with the following displacements for any overlying dislocation:

\begin{equation}
\mathbf{d}_{5x} = \acute{\mathbf{d}}_{5x} \quad d_{5x} = \frac{a}{2}\frac{|3\gamma^{2}-7|}{\sqrt{\gamma^{2}+1}}
\end{equation}

\begin{equation}
\mathbf{d}_{5z}=\mathbf{d}_{2z}
\end{equation}

The geometrical implications of the theory are equivalent to basic trigonometry relative to the dark blue triangle in Figure~\ref{Pmaster}.

\subsection{Case of $\boldsymbol{\acute{\eta}}_{2/6}=2/3\hkl[-2 -2 4 -3]$}
\label{d6}

Substituting $\boldsymbol{\acute{\eta}}_{2/6}=\hkl[-2 -2 4 -3]$ and $\mathbf{f}=1/6\hkl[2 0 -2 -3]$ in Equations~\ref{shearf}-\ref{shufflefz}, we obtain:

\begin{equation}
\acute{\mathbf{b}}_{6} = \frac{2\gamma^2-4}{3(\gamma^{2}+1)}[\bar{1}\bar{1}23]
\end{equation}

\begin{equation}
\acute{\mathbf{d}}_{6x} = \frac{9-3\gamma^2}{6(\gamma^{2}+1)}[11\bar{2}\bar{3}]
\end{equation}

Therefore, $\lVert\mathbf{\acute{b}}_6\rVert > \lVert\mathbf{\acute{d}}_{6x}\rVert$, and as such, no twinning dislocation can glide on this plane which will instead be acted on by shuffles, for any overlying dislocation, in the $\mathbf{x}$, and $\mathbf{z}$ directions with the following displacements:

\begin{equation}
\mathbf{d}_{6x} = \acute{\mathbf{d}}_{6x}, \quad d_{6x} = \frac{a}{2}\frac{|9-3\gamma^2|}{\sqrt{\gamma^{2}+1}}
\end{equation}

\begin{equation}
\mathbf{d}_{6z}= \mathbf{d}_{5z}= \mathbf{d}_{2z}
\end{equation}

The geometrical implications of the theory simplify to trigonometry applied to the dark orange triangle in Figure~\ref{Pmaster}.

\subsection{Case of $\boldsymbol{\acute{\eta}}_{2/7}=1/3\hkl[-5 -5 10 -6]$}
\label{B7}

\begin{figure}[h]
\centering
\includegraphics[width = 0.7\textwidth]{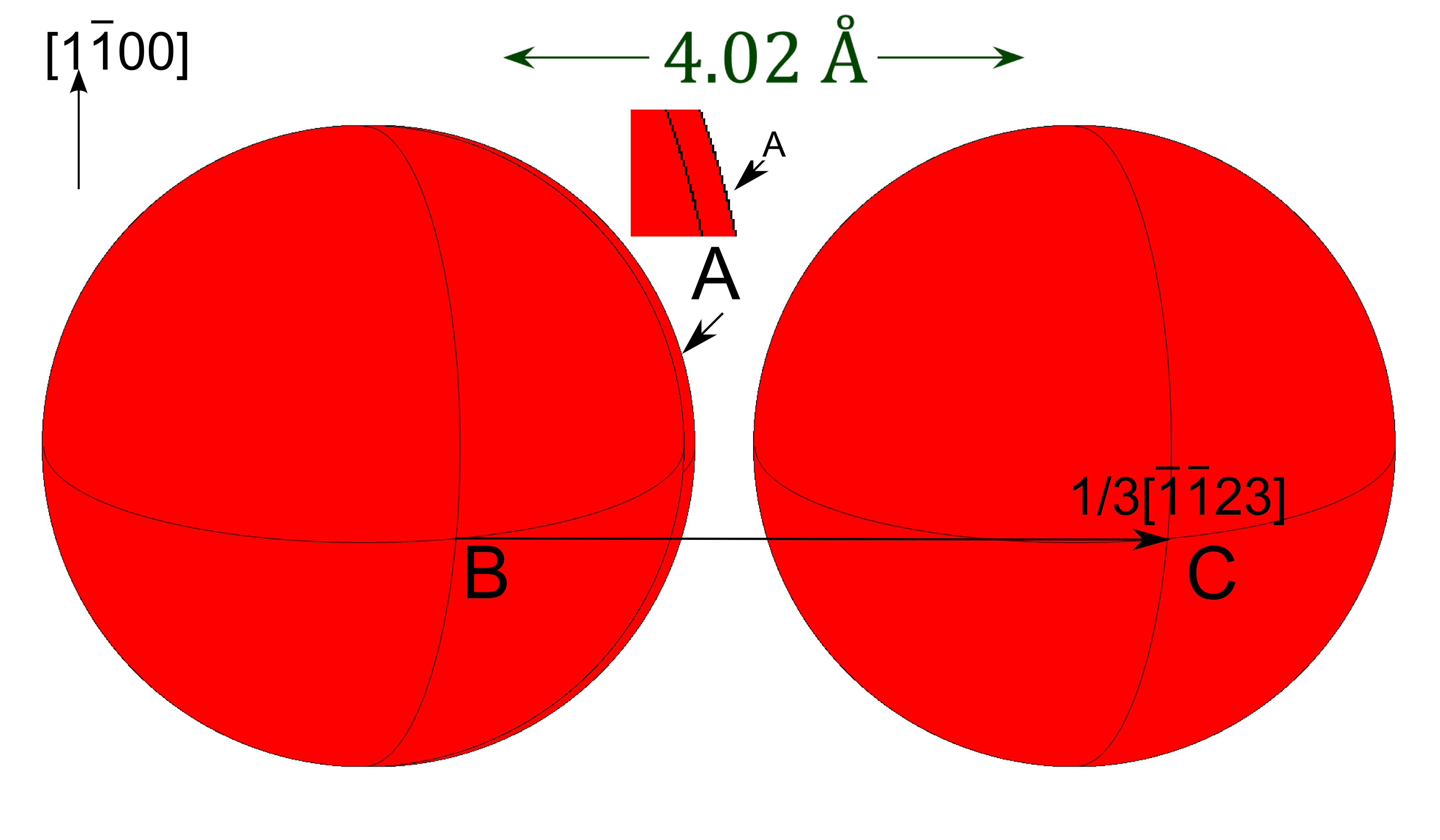}
	
\caption{\hkl(1 1 -2 2) projection of the $\mathbf{b}_7$ vector showing that atom A and atom B have nearly identical projections, and there is no out of shear plane component of the dislocation.}
\label{twinview}
\end{figure}

Substituting $\boldsymbol{\acute{\eta}}_{2/7}=1/3\hkl[-5 -5 10 -6]$ and $\mathbf{f}=1/6\hkl[2 0 -2 -3]$ in Equations~\ref{shearf}-\ref{shufflefz}, we obtain 

\begin{equation}
\acute{\mathbf{b}}_{7}=\frac{2\gamma^{2}-5}{3(\gamma^{2}+1)}[\bar{1}\bar{1}23] 
\end{equation}

\begin{equation}
\acute{\mathbf{d}}_{7x}=\frac{11-3\gamma^{2}}{6(\gamma^{2}+1)}[11\bar{2}\bar{3}] 
\end{equation}

Therefore, $\lVert\mathbf{\acute{b}}_7\rVert << \lVert\mathbf{\acute{d}}_{7x}\rVert$, and as such, $\mathbf{b}_7$ with $\boldsymbol{\eta}_{2/7}$=1/3\hkl[-5 -5 10 -6], and $K_{2/7}=\hkl(1 1 -2 -5)$, is a possible twinning dislocation candidate accommodating compression of the \hkl<c>-axis for HCP metals with $\gamma>\sqrt{5/2}$ (almost all HCP metals) and is a possible twinning dislocation candidate accommodating tension of the \hkl<c>-axis for HCP metals with $\gamma<\sqrt{5/2}$ (Be). The Burgers vector is given from above by:

\begin{equation}
\mathbf{b}_{7}=\acute{\mathbf{b}}_{7}, \quad b_{7} = a\frac{2\gamma^{2}-5}{\sqrt{\gamma^{2}+1}}
\end{equation}

\begin{table}[h]
\caption{$\hkl{11-22}$ shuffle and shear components for  $\mathbf{b}_1$,$\mathbf{b}_3$, $\mathbf{b}_4$, and $\mathbf{b}_7$ twinning dislocation candidates}
\renewcommand{\arraystretch}{1.5}
\centering
\begin{tabular}{c c c c}
\toprule
Plane & Shuffle displacement & Shear displacement & Net shuffle\\
$n$ & $d_{nx}$ & $d_s$ & $\Delta_{nx}^t$\\
\midrule
\multicolumn{4}{c}{$\mathbf{b_1}$ ($t=1$)}\\
1 & $\frac{1}{3(\gamma^2+1)}[11\bar{2}\bar{3}]$ & $\frac{1}{3(\gamma^2+1)}[11\bar{2}\bar{3}]$ & 0\\
\multicolumn{4}{c}{$\mathbf{b_3}$ ($t=3$)}\\
1 & $\frac{1}{3(\gamma^2+1)}[11\bar{2}\bar{3}]$ & $\frac{1}{3}\mathbf{b_3}$ & $\frac{1}{9}[11\bar{2}\bar{3}]$ \\
2 & $\frac{3-\gamma^2}{6(\gamma^2+1)}[11\bar{2}\bar{3}]$ & $\frac{2}{3}\mathbf{b_3}$ & $\frac{1}{18}[11\bar{2}\bar{3}]$ \\
3 & $\frac{\gamma^2-2}{3(\gamma^2+1)}[\bar{1}\bar{1}23]$ & $\mathbf{b_3}$ & 0 \\
\multicolumn{4}{c}{$\mathbf{b_4}$ ($t=4$)}\\
1 & $\frac{1}{3(\gamma^2+1)}[11\bar{2}\bar{3}]$ & $\frac{1}{4}\mathbf{b_4}$ & $\frac{1}{12}[11\bar{2}\bar{3}]$ \\
2 & $\frac{3-\gamma^2}{6(\gamma^2+1)}[11\bar{2}\bar{3}]$ & $\frac{1}{2}\mathbf{b_4}$ & 0 \\
3 & $\frac{\gamma^2-2}{3(\gamma^2+1)}[\bar{1}\bar{1}23]$ & $\frac{3}{4}\mathbf{b_4}$ & $\frac{1}{12}[\bar1\bar{1}23]$ \\
4 & $\frac{3-\gamma^2}{3(\gamma^2+1)}[11\bar{2}\bar{3}]$ & $\mathbf{b_4}$ & 0 \\
\multicolumn{4}{c}{$\mathbf{b_7}$ ($t=7$)} \\
1 & $\frac{1}{3(\gamma^2+1)}[11\bar{2}\bar{3}]$ & $\frac{1}{7}\mathbf{b_7}$ & $\frac{2}{21}[11\bar{2}\bar{3}]$ \\
2 & $\frac{3-\gamma^2}{6(\gamma^2+1)}[11\bar{2}\bar{3}]$ & $\frac{2}{7}\mathbf{b_7}$ & $\frac{1}{42}[11\bar{2}\bar{3}]$ \\
3 & $\frac{\gamma^2-2}{3(\gamma^2+1)}[\bar{1}\bar{1}23]$ & $\frac{3}{7}\mathbf{b_7}$ & $\frac{1}{21}[\bar1\bar{1}23]$ \\
4 & $\frac{3-\gamma^2}{3(\gamma^2+1)}[11\bar{2}\bar{3}]$ & $\frac{4}{7}\mathbf{b_7}$ & $\frac{1}{21}[11\bar{2}\bar{3}]$ \\
5 & $\frac{3\gamma^2-7}{6(\gamma^2+1)}[\bar{1}\bar{1}23]$ & $\frac{5}{7}\mathbf{b_7}$ & $\frac{1}{42}[\bar1\bar{1}23]$\\
6 & $\frac{9-3\gamma^2}{6(\gamma^2+1)}[11\bar{2}\bar{3}]$ & $\frac{6}{7}\mathbf{b_7}$ & $\frac{1}{14}[11\bar2\bar3]$\\
7 & $\frac{2\gamma^2-5}{3(\gamma^2+1)}[\bar{1}\bar{1}23]$ & $\mathbf{b_7}$ & 0 \\
\end{tabular}
\label{Ptable}
\end{table}	

This $\mathbf{b}_{7}$ dislocation has not been previously mentioned or studied in the literature as a possible twinning dislocation. This dislocation candidate corresponds to an extremely small ``deviation'' from the exact stacking sequence of \hkl{1 1 -2 2} which practically vanishes for $\gamma$ values close to that of Hf ($\gamma=1.581$ is approximately $\sqrt{5/2}=1.5811$). This deviation is better illustrated by the \hkl(1 1 -2 2) projection in Figure~\ref{twinview}, which shows \hkl(1 1 -2 2) projection of the two \hkl{1 1 -2 2} planes situated at $-7^{th}$ and $7^{th}$ planes (with respect to the $0^{th}$ plane highlighted with dashed lines in Figure~\ref{Pmaster}).  Atoms B and C from one \hkl{1 1 -2 2} plane are shown to illustrate the spacing in the  $\mathbf{x}$ direction.  Atom A from a second \hkl{1 1 -2 2} plane is included to show the nearly identical positions of Atoms A and B with respect to the spacing of Atoms B and C.

All these twinning dislocations rotate the basal pole by the same angle, $\alpha$, required for mirror symmetry on the \hkl{1 1 -2 2}, which can be derived using any of the above triangles:

\begin{equation}
\alpha= 2 \arctan\left(\frac{1}{\gamma}\right)
\end{equation}

\subsection{Net shuffles}
For flat planes, Equation~\ref{netshuffle} reduces to:

\begin{equation}
\mathbf{\Delta}^{t}_{nx}=\mathbf{d}_{nx}-\frac{n}{t}\mathbf{b}_t
\label{netshuffleflat}
\end{equation}

Applying this equation for all above twinning dislocation candidates $\mathbf{b}_1$,$\mathbf{b}_3$, $\mathbf{b}_4$, and $\mathbf{b}_7$, the net shuffle vectors; $\mathbf{\Delta}^{m}_{n}$, can be extracted for $n=1,...,7$ and $t=1,3,4,7$. These vectors are summarized in Table~\ref{Ptable}.  Interestingly, the net shuffles required for every twinning dislocation candidate do not depend on the c/a ratio $\gamma$.  Instead, in applying Equation \ref{netshuffleflat}, the magnitude of the shuffle and the partial shear from the twinning dislocation sum to produce a multiple of $\gamma^2+1$ in the numerator, which cancels out this term in the denominator.  Thus, all net shuffles are rational.  Since the twinning dislocations are dependent on $\gamma$, this may help to explain the preference of certain twinning dislocation candidates for various HCP metals or as a function of temperature, etc.

\section{Candidacy in \hkl{10-12} twinning}

Our analyses of the stacking sequence of \hkl{1 0 -1 2} planes indicate that if a very small ``deviation'' exists, then it lies at a stacking distance greater than 40. This is a relatively high value which corresponds to a plane which is too distant to play a significant role in the deformation caused by shear on the \hkl{1 0 -1 2} planes. 

The heights $h$ and $h_m$ of \hkl{10-12} planes being equal to $ h= a\gamma\sqrt{3}/(2\sqrt{\gamma^2 + 3})$ and $h_m=a\gamma\sqrt{3}/(6\sqrt{\gamma^2+3})$, respectively, the first four sets of possibilities ($n=1-4$) for the three $\boldsymbol{\acute{\eta}}_{2/n_0,n_a,n_b}$ vectors satisfying Equation~\ref{condition} are reported in Table~\ref{eta2TT}. For each of these vectors, a dedicated analysis follows where the Burgers vectors and shuffles are computed through the systematic use of Equations~\ref{shearc0}-\ref{shufflecy}. The vector $\mathbf{f}$ for the \hkl{10-12} is $1/3$\hkl[-12-10] and thus, $\mathbf{f} \cdot \boldsymbol{\eta}_1$ vanishes. Figure~\ref{Tmaster} which depicts the \hkl{-1 2 -1 0} projection of 11 stackings of \hkl{1 0 -1 2} planes illustrates some of the possibilities of $\boldsymbol{\acute{\eta}}_{2/n0,na,nb}$ and their corresponding required shear and shuffle displacements to construct the twin lattice with respect to a randomly selected mirror plane denoted by ``$0$'' (final position of the twin boundary).

\begin{table}[ht]
\caption{ $\boldsymbol{\acute{\eta}}_{2/n}$ interrogation results}
\centering
\begin{tabular}{c c c c}
\addlinespace
\toprule
Plane & $\boldsymbol{\acute{\eta}}_{2/n_0}$ & $\boldsymbol{\acute{\eta}}_{2/n_a}$ & $\boldsymbol{\acute{\eta}}_{2/n_b}$\\[1ex]
\midrule
$1$ & $\hkl[0 0 0 1]$ & $1/6\hkl[2 0 -2 3]$ & $1/6\hkl[4 0 -4 3]$\\
$2$ & $\hkl[1 0 -1 1]$ & $1/6\hkl[8 0 -8 3]$ & $1/6\hkl[4 0 -4 9]$\\
$3$ & $\hkl[1 0 -1 2]$ & $1/6\hkl[8 0 -8 9]$ & $1/6\hkl[10 0 -10 9]$\\
$4$ & $\hkl[-2 0 -2 2]$ & $1/6\hkl[8 0 -8 15]$ & $1/6\hkl[10 0 -10 15]$ \\[1ex]

\bottomrule
\end{tabular}
\label{eta2TT}
\end{table}

\subsection{Case of $\boldsymbol{\acute{\eta}}_{2/1_0}=\hkl[0 0 0 1]$ }

Substituting $\boldsymbol{\acute{\eta}}_{2/1_0}=\hkl[0 0 0 1]$, $\boldsymbol{\acute{\eta}}_{2/1_a}=1/6\hkl[2 0 -2 3]$, and $\boldsymbol{\acute{\eta}}_{2/1_b}=1/6\hkl[4 0 -4 3]$ into Equations~\ref{shearc0}-\ref{shufflecy} we obtain:

\begin{equation}
\mathbf{\acute{b}}_{1_0}=\frac{\gamma^2}{\gamma^2+3}\hkl[-1011]
\end{equation}

\begin{equation}
\mathbf{\acute{b}}_{1_a} = \frac{\gamma^2-2}{2(\gamma^2+3)}[10\bar1\bar1]
\end{equation}

\begin{equation}
\mathbf{\acute{b}}_{1_b}=\frac{4-\gamma^2}{2(\gamma^2+3)}[\bar1011]
\end{equation} 

\begin{figure}[ht]
\centering
\includegraphics[width = \textwidth]{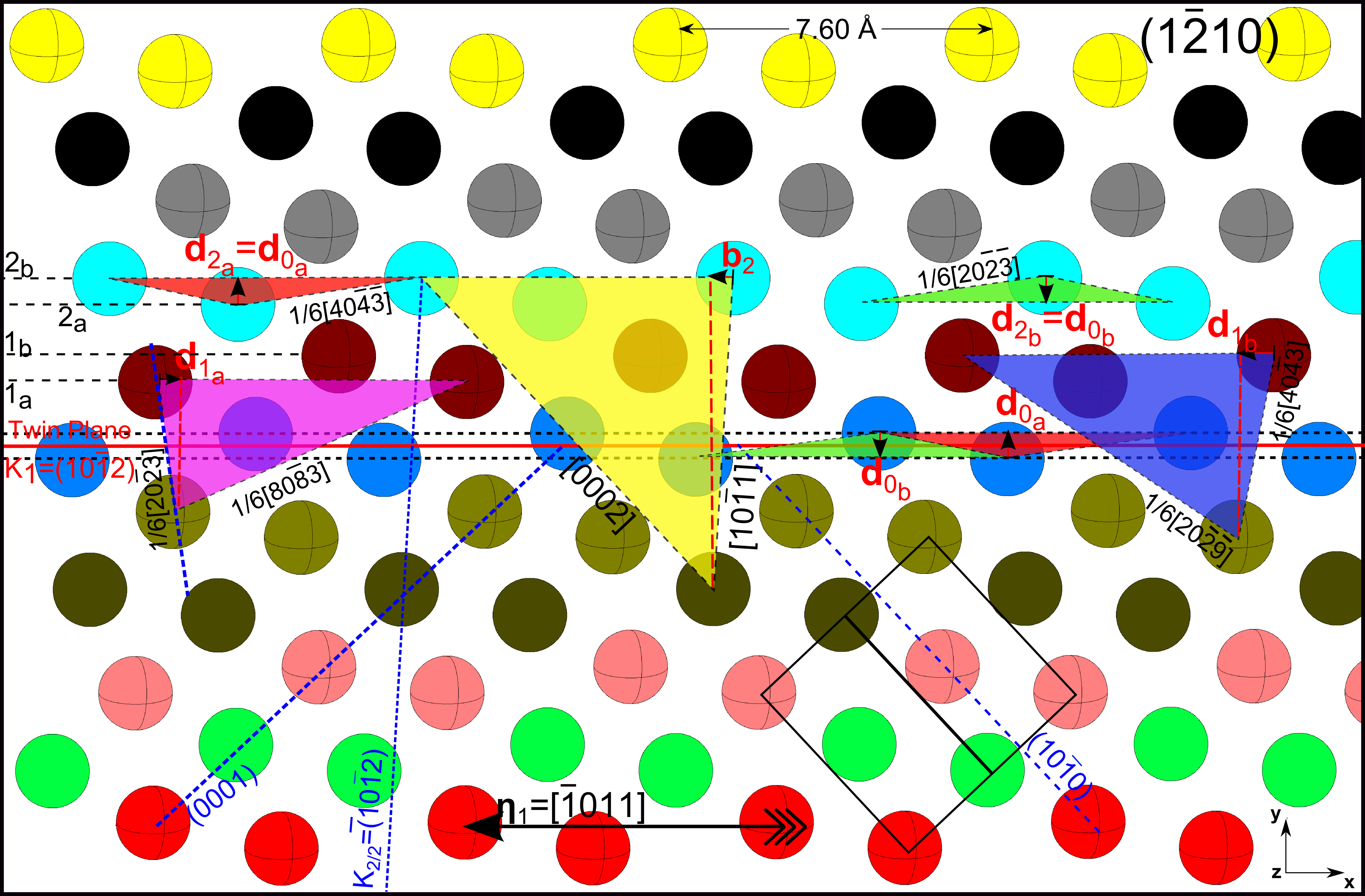}
\caption{\hkl(-1 2 -1 0) projection of atom positions in pure Mg constructed by CrystalMaker\textsuperscript{\textregistered} \citep{CrystalMaker} displaying 11 layers of \hkl(10-12) planes revealing the crystallography of shear and shuffle of \hkl(1 0 -1 2)\hkl[-1 0 1 1] tensile twinning  through the theory outlined in the beginning of this paper. Without loss of generality, the final position for each twinning dislocation candidate is assumed to be at the $0^{th}$ plane which is the sixth plane counting from the bottom plane.}
\label{Tmaster}
\end{figure}

Therefore, $\lVert\mathbf{\acute{b}}_{1_0}\rVert \geq \lVert\mathbf{\acute{b}}_{1_a}\rVert$ and as such, no twinning dislocation could operate on this plane. Instead two different and opposite shuffles in the $\mathbf{x}$ direction must operate on the sub-planes $1_a$ and $1_b$ for any twinning dislocation $\mathbf{b}_t$ such that $t>n$. These shuffles are $\mathbf{d}_{1_ax}=\mathbf{\acute{b}}_{1_a}$ and $\mathbf{d}_{1_bx}=\mathbf{\acute{b}}_{1_b}$.  The geometrical implications of the theory simplify to trigonometry applied to the dark pink and blue triangles highlighted in Figure~\ref{Tmaster}.

\subsection{Case of $\boldsymbol{\acute{\eta}}_{2/2_0}=\hkl[1 0 -1 1]$ }

Substituting $\boldsymbol{\acute{\eta}}_{2/2_0}=\hkl[1 0 -1 1]$, $\boldsymbol{\acute{\eta}}_{2/2_a}=1/6\hkl[8 0 -8 3]$, and $\boldsymbol{\acute{\eta}}_{2/2_b}=1/6\hkl[4 0 -4 9]$ into Equations~\ref{shearc0}-\ref{shufflecy} we obtain:

\begin{equation}
\mathbf{\acute{b}}_{2_0}=\frac{3-\gamma^2}{\gamma^2+3}[\bar1011]
\end{equation}

\begin{equation}
\mathbf{\acute{b}}_{2_a} = \frac{8-\gamma^2}{2(\gamma^2+3)}[10\bar1\bar1]
\end{equation}

\begin{equation}
\mathbf{\acute{b}}_{2_b}=\frac{4-3\gamma^2}{2(\gamma^2+3)}[\bar1011]
\end{equation}

Therefore, $\lVert\mathbf{\acute{b}}_{2_0}\rVert < \lVert\mathbf{\acute{b}}_{2_a}\rVert$, and as such, $\mathbf{b}_2 = \mathbf{\acute{b}}_{2_0}$ is a twinning dislocation candidate. In addition to $\mathbf{x}$ shuffles in the the first plane as specified above, this twinning dislocation must be accompanied by alternating $\mathbf{y}$ shuffles which are equal to those in $K_1$:

\begin{equation}
\mathbf{d}_{2_ay}=\mathbf{d}_{0_ay}=\mathbf{\acute{d}}_{2_ay}= \frac{a\gamma\sqrt{3}}{6\sqrt{\gamma^2 + 3}}\mathbf{m}
\label{d0ay}
\end{equation}

\begin{equation}
\mathbf{d}_{2_by}=\mathbf{d}_{0_by}=\mathbf{\acute{d}}_{2_by}= -\frac{a\gamma\sqrt{3}}{6\sqrt{\gamma^2 + 3}}\mathbf{m}
\label{d0by}
\end{equation}

For this dislocation $\boldsymbol{\eta}_2=\boldsymbol{\acute{\eta}}_{2_0}=\hkl[1 0 -1 1]$, and $K_2=\hkl(1 0 -1 -2)$. For the \hkl<c>-axis, this dislocation must accommodate tension for $\gamma < \sqrt{3}$ and compression for $\gamma > \sqrt{3}$. That is, we recover the active form of the classical $\mathbf{b}_2$ which was previously derived by \citet{Serra1988,Serra1986,Christian1995}. However, these authors did not suggest a combined theory that yields both the Burgers vectors of the twinning dislocation and the shuffles.  The geometrical implications of the theory simplify to basic trigonometry performed on the yellow, green, and dark orange triangles in Figure~\ref{Tmaster}.

\subsection{Case of $\boldsymbol{\acute{\eta}}_{2/3_0}=\hkl[1 0 -1 2]$ }

Substituting $\boldsymbol{\acute{\eta}}_{2/3_0}=\hkl[1 0 -1 2]$, $\boldsymbol{\acute{\eta}}_{2/2_a}=1/6\hkl[8 0 -8 9]$, and $\boldsymbol{\acute{\eta}}_{2/2_b}=\hkl[10 0 -10 9]$ into Equations~\ref{shearc0}-\ref{shufflecy} we obtain:

\begin{equation}
\mathbf{\acute{b}}_{3_0}=\frac{3-2\gamma^2}{\gamma^2+3}[\bar1011]
\end{equation}

\begin{equation}
\mathbf{\acute{b}}_{3_a} = \frac{8-3\gamma^2}{2(\gamma^2+3)}[10\bar1\bar1]
\end{equation}

\begin{equation}
\mathbf{\acute{b}}_{3_b}=\frac{10-3\gamma^2}{2(\gamma^2+3)}[\bar1011]
\end{equation}

Therefore, $\lVert\mathbf{\acute{b}}_{3_0}\rVert \geq \lVert\mathbf{\acute{b}}_{3_a}\rVert$ and as such, no twinning dislocation could operate on this plane. Instead two different shuffles in the $\mathbf{x}$ direction must operate on the sub-planes $3_a$ and $3_b$ for any twinning dislocation $\mathbf{b}_t$ such that $t>n$. These shuffles are $\mathbf{d}_{3_ax}=\mathbf{\acute{b}}_{3_a}$ and $\mathbf{d}_{3_bx}=\mathbf{\acute{b}}_{3_b}$.

\subsection{Case of $\boldsymbol{\acute{\eta}}_{2/4_0}=\hkl[-2 0 -2 2]$ }

Substituting $\boldsymbol{\acute{\eta}}_{2/4_0}=\hkl[-2 0 -2 2]$, $\boldsymbol{\acute{\eta}}_{2/4_a}=1/6\hkl[8 0 -8 15]$, and $\boldsymbol{\acute{\eta}}_{2/4_b}=1/6\hkl[10 0 -10 15]$ into Equations~\ref{shearc0}-\ref{shufflecy} we obtain:

\begin{equation}
\mathbf{\acute{b}}_{4_0}=\frac{2(3-\gamma^2)}{\gamma^2+3}[\bar1011]
\end{equation}

\begin{equation}
\mathbf{\acute{b}}_{4_a} = \frac{8-5\gamma^2}{2(\gamma^2+3)}[10\bar1\bar1]
\end{equation}

\begin{equation}
\mathbf{\acute{b}}_{4_b}=\frac{10-5\gamma^2}{2(\gamma^2+3)}[\bar1011]
\end{equation}

Therefore, $\lVert\mathbf{\acute{b}}_{4_0}\rVert < \lVert\mathbf{\acute{b}}_{4_a}\rVert$, and as such, $\mathbf{b}_4 = 2\mathbf{\acute{b}}_{2_0}$ is a twinning dislocation candidate. In addition to $\mathbf{x}$ shuffles on the first and third planes as specified above, this twinning dislocation must be accompanied by $\mathbf{y}$ shuffles equal to those on the second plane and on $K_1$. For the \hkl<c>-axis, this dislocation must accommodate tension for $\gamma < \sqrt{3}$ and compression for $\gamma > \sqrt{3}$. It has $\boldsymbol{\eta}_2=\boldsymbol{\acute{\eta}}_{2_0}=\hkl[-2 0 -2 2]$, and $K_2=\hkl(-1 0 1 2)$. Observe that the Burgers vector of this dislocation is twice that of $\mathbf{b}_2$ despite the larger step height. That is, no dislocation candidate could operate on \hkl{10-12} with a step height larger than two interplanar spacing. Furthermore, as no twinning dislocation could glide on the first plane, $\mathbf{b}_2$ is the only twinning dislocation on \hkl{10-12} planes having no candidate to compete with.  Interestingly, the following analysis for $\boldsymbol{\acute{\eta}}_{2/4_0}=\hkl[-2 0 -2 2]$ is also easily predicted by considering that $\hkl[-2 0 -2 2]=2\hkl[-1 0 -1 1]=2\boldsymbol{\acute{\eta}}_{2/2_0}$, which results in $\mathbf{b}_4=2\mathbf{b}_2$.

\subsection{Net Shuffles}

\begin{table}[ht]
\caption{\hkl{10-12} shuffle displacements, shear displacements, and net shuffles for $\mathbf{b}_2$ twinning dislocation}
\centering
\renewcommand{\arraystretch}{1.5}
\begin{tabular}{c c c c}
\toprule
Plane & Shuffle displacement & Shear displacement & Net Shuffle\\
n & $\mathbf{d}_n$ & $\mathbf{d}_s$ & $\Delta^t_n$\\
\midrule
$0_a$ & $\frac{a\gamma\sqrt{3}}{6\sqrt{\gamma^2 + 3}}\mathbf{m}$ & $\frac{1}{12}\mathbf{b}_2$ & $\frac{1}{12}[\bar101\bar1]$ \\
$0_b$ & $-\frac{a\gamma\sqrt{3}}{6\sqrt{\gamma^2 + 3}}\mathbf{m}$ & $-\frac{1}{12}\mathbf{b}_2$ & $\frac{1}{12}[10\bar11]$ \\
$1_a$ & $\frac{4-\gamma^2}{2(\gamma^2+3)}[\bar1011]$ & $\frac{7}{12}\mathbf{b}_2$ & $\frac{1}{12}[\bar1011]$ \\
$1_b$ & $\frac{\gamma^2-2}{2(\gamma^2+3)}[10\bar1\bar1]$ & $\frac{5}{12}\mathbf{b}_2$ & $\frac{1}{12}[10\bar1\bar1]$ \\
\bottomrule
\end{tabular}
\label{Ttable}
\end{table}

Applying Equation~\ref{netshuffle} for the unique $\mathbf{b}_2$, the net shuffle values $\mathbf{\Delta}^{t}_{n}$ for the subplanes $0_a$, $0_b$, $1_a$, and $1_b$ are reported in Table \ref{Ttable}.  Similar to Table~\ref{Ptable}, the net shuffles required for the $\mathbf{b}_2$ twinning dislocation candidate do not depend on the c/a ratio $\gamma$.  Again, all net shuffles on the 0$^{th}$ and 1$^{st}$ planes are found to be rational. 

\section{Discussions}

This paper describes a generalized crystallographic framework that predicts the analytical expressions of both shuffle and shear candidates for any given compound twin mode in HCP metals. The corresponding formulations were derived based on the hypothesis that at any plane above the final position of the composition plane, shear and shuffle would compete to bring parent atoms to the correct twin positions. The theory can predict all candidates of twinning dislocation Burgers vectors and twinning shuffles. Then, through a unifying minimum displacement criterion, it adjudicates whether shear or ``only'' shuffle on a specific plane should be favorable.

\begin{figure}[ht]
\centering
\subfloat{\includegraphics[width = 0.45\textwidth]{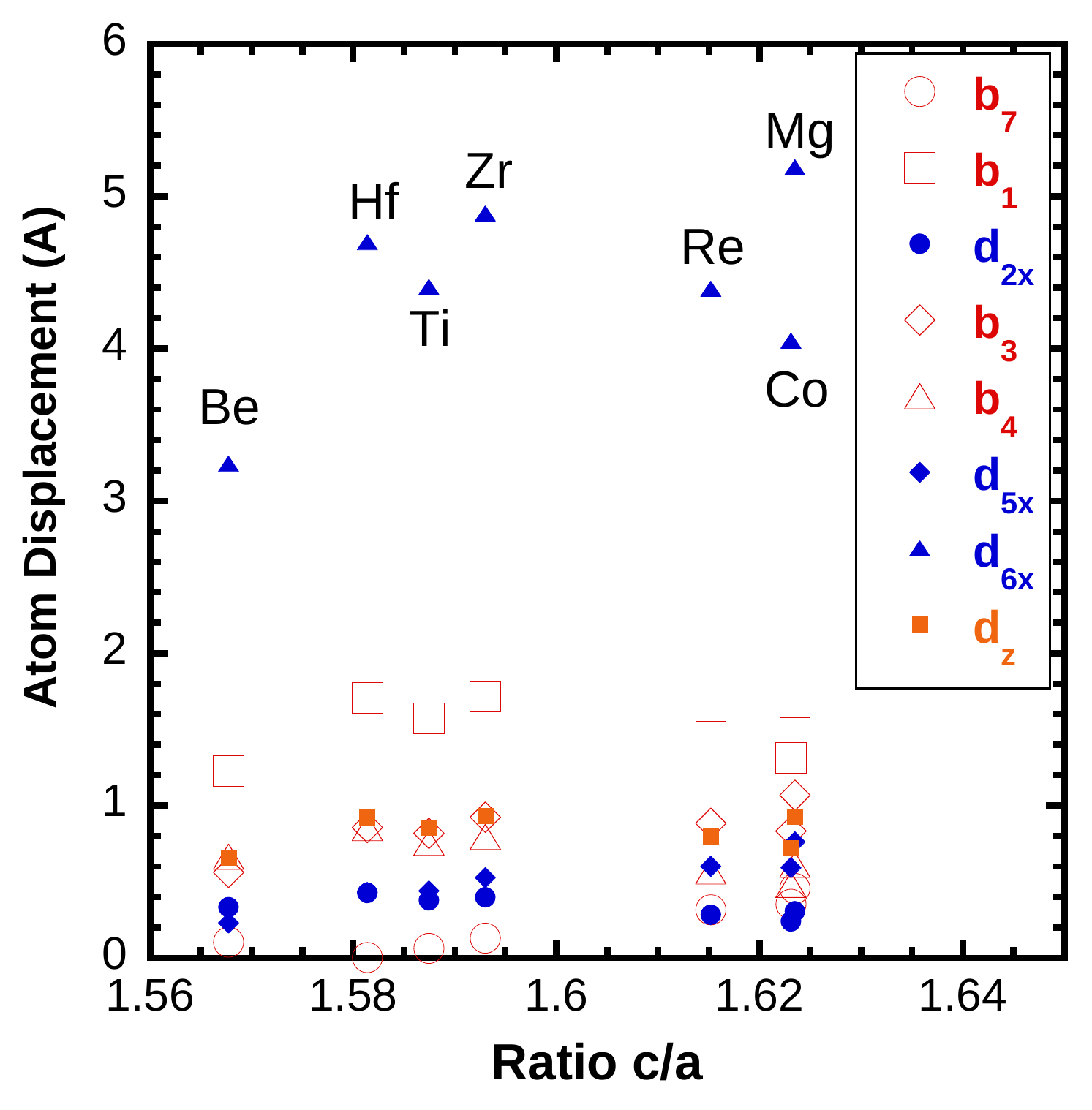}\label{rsa}}
\quad
\subfloat{\includegraphics[width = 0.46\textwidth]{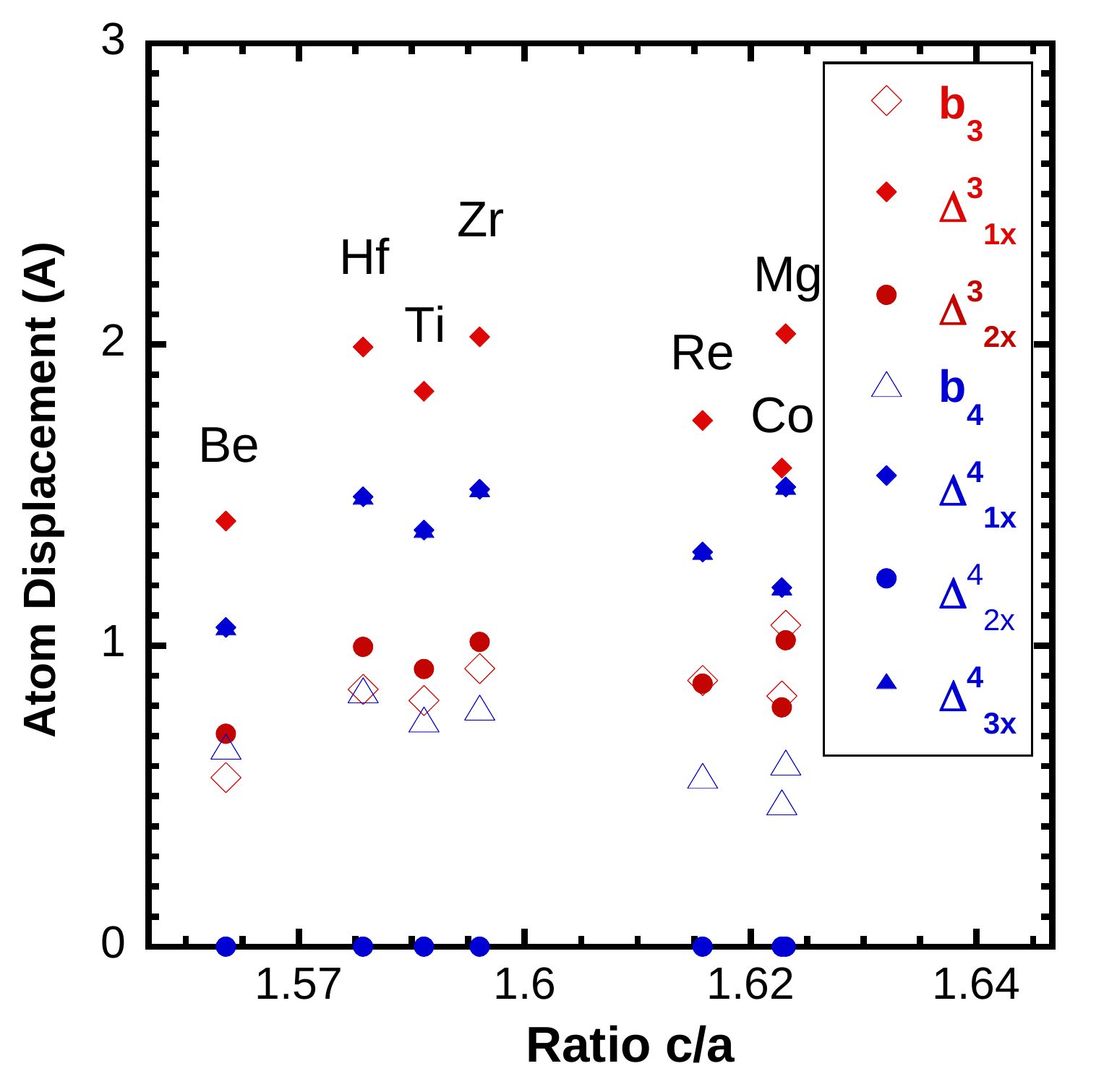}\label{rsb}}
\caption{(a) $\mathbf{x}$ Shuffle displacements and Burgers vector magnitudes of each \hkl{1 1 -2 2} twinning dislocation candidate in Be, Hf, Ti, Zr, Re, Co, and Mg. Observe that the \hkl{1 1 -2 2} stacking sequence ``deviation'' $\mathbf{b}_7$ vanishes for Hf, and (b) Net $\mathbf{x}$ shuffle magnitudes for $\mathbf{b}_3$ and $\mathbf{b}_4$ twinning dislocation candidates in  HCP metals. Observe that  $\boldsymbol{\Delta}^{4}_{2x}$ vanishes for all HCP metals, and that $\boldsymbol{\Delta}^{4}_{1x}=\boldsymbol{\Delta}^{4}_{3x}$.}
\end{figure}

The theory was tested for flat \hkl{11-22} and corrugated \hkl{10-12} twinning mode planes. For example, Figure~\ref{rsa} shows the absolute values of all shuffle displacement and Burgers vector candidates for \hkl{11-22} twinning for HCP metals with $c/a$ ratios smaller than the perfect $c/a$ ratio. The theory not only recovered the correct expression of all twinning dislocation candidates which were previously identified from the admissible interfacial defect theory by \citet{Pond1983,Pond1983a,Serra1988,Serra1991}, but it simultaneously yielded the shuffle displacement and net shuffle vectors required by the twinning process. This is a key feature of the theory as shuffle magnitudes and directions have not been previously derived. Analytical expressions of shuffles are essential for quantifying the mechanisms by which they affect the atomic structure of the twinning dislocation core and, thus, their mobility \citep{Serra1988,Serra1991}. Moreover, the theory yielded a new twinning dislocation candidate for \hkl{11-22}, $\mathbf{b}_7$, twinning which was not previously reported. In fact, the theory highlights many crystallographic characteristics of twinning in HCP structures. For instance, despite the fact that pyramidal planes in HCP structures are non-merohedral, i.e. have no stacking sequence, the smallest deviation from the exact stacking sequence is exactly repeatable with an order, say $p$. Thus, the interrogation process of $\boldsymbol{\acute{\eta}}_{2/n}$ stops at $n=p$ when $\boldsymbol{\acute{\eta}}_{2/p}=\boldsymbol{\acute{\eta}}_{2/2p}$. As an example, for \hkl{11-22} twinning, $p=7$ and thus $\mathbf{b}_7$ was the smallest possible deviation from an exact stacking sequence and thus the smallest possible Burgers vector. Any plane above will have its possible corresponding shear or shuffle displacements an integer multiple of those at some plane which is seven interplanar spacings below. For \hkl{10-12} planes, $p=2$, and at the fourth plane, the stacking sequence deviation is exactly double that of $\mathbf{b}_2$.

Furthermore, it was found that the twinning modes which have the vector $\mathbf{f}$ normal to the shear plane do not require shuffles outside the shear plane. This is interestingly the case of all observed twin modes with \hkl{11-20} as the shear plane, i.e. having the shear plane with an $ABAB$ stacking sequence. Moreover, all these twin modes have $K_1$ corrugated, and they all required $\mathbf{y}$-shuffles in the shear plane at every other plane starting from $K_1$. 

When the vector $\mathbf{f}$ is not normal to the shear plane, $\mathbf{z}$-shuffles normal to the plane of shear are binding when the step height of a twinning dislocation candidate is greater than a single interplanar spacing. This is found to be the case for all observed twin modes having \hkl{10-10} ($ABCDABCD$ stacking sequence) as the shear plane. Moreover, all these observed twin modes have flat $K_1$. The theory predicts exactly on which planes these $\mathbf{z}$-shuffles are required. For instance, for \hkl{11-22} $\mathbf{b}_7$, the second, fifth and sixth planes must undergo both $\mathbf{z}$ and $\mathbf{x}$ shuffles.

The introduction of the shuffle displacements was necessary for the analytical derivation of the net shuffles. The partition between net shuffles and shear displacement on lower planes will enable the partition between deviatoric stress and hydrostatic pressure to be clearly made in case the energy consumed by twinning dislocation passage is to be calculated. In fact, as shuffles in principal bring no shape change, they should be correlated to hydrostatic pressure, and not to the deviatoric stress. This partition is by convention in diffusion practice where $\mathbf{\sigma}_{kk}$ is the only stress component driver of solid diffusion of species \citep{Elkadiri2008}. 

It is interesting to observe that in \hkl{11-22} twinning, while net $\mathbf{x}$ shuffles required by $\mathbf{b}_3$ at the second plane are higher than the corresponding shuffle displacements, they exactly vanish for $\mathbf{b}_4$ dislocation (Figure~\ref{rsb}). Furthermore, net $\mathbf{x}$  shuffles for $\mathbf{b}_4$ dislocation at the first and second plane are exactly the same for all HCP metals. These two phenomena are rather intriguing and may provide an explanation why \citet{Serra1988} computed a wider core for $\mathbf{b}_4$ than for $\mathbf{b}_3$, although the latter is the observed twinning dislocation in Ti and Zr. Apart from $\mathbf{z}$ shuffles, the $\mathbf{b}_4$ shear brings atoms at the second plane to their exact twin position and is better than $\mathbf{b}_3$ in bringing atoms on the first plane to their twin positions. However, as $\mathbf{b}_4$ accommodates tension of the \hkl<c>-axis for Ti and Zr, it has to compete with \hkl{10-12} (and \hkl{11-21} as well), which has a much more mobile twinning dislocation. Hence, it is unlikely that $\mathbf{b}_4$ nucleates in HCP metals for which it accommodates \hkl<c>-tension (i.e. $c/a < \sqrt{3}$). Furthermore, in metals for which $\mathbf{b}_4$ accommodates \hkl<c>-axis compression, namely Zn and Cd, this dislocation candidate must again compete with \hkl{10-12} twinning, as \hkl{10-12} twinning reverses the shear sign at a value of $c/a=\sqrt{3}$ exactly like $\mathbf{b}_4$ (meaning both accommodate \hkl<c>-axis compression). $\mathbf{b}_3$ is much more favored than $\mathbf{b}_4$ in that the shear reversal occurs only at $\sqrt{2}$, below which no metallic $c/a$ ratio exists.  In \hkl{10-12} twinning, the shuffle displacements are considerably simpler and of lower magnitude than those involved in \hkl{11-22} twinning. These shuffles are required only to change the sense of corrugation so that prismatic planes become basal planes and vice-versa. This is why $\mathbf{y}$ shuffles are, however, required in \hkl{10-12} and not in \hkl{11-22}. Nonetheless, this is compensated by the absence of shuffles out of the shear plane.

In the above calculations, it was necessary to consider the twin plane as the bisector of the $0_a$ and $0_b$ sub-planes; otherwise, it would be difficult to justify such an unphysical bias. The plane is however, subject to $\mathbf{y}$ in opposite directions across the composition plane. Thus, the type of shuffles identified by the theory would give alternating stress sign from one atom to the other atom in $K_1$. This is in agreement with the prior simulation results \citep{Serra1988,Serra1991}. However, it is possible to imagine that these $\mathbf{y}$ shuffles at the $K_1$ plane are not completed until the interface has moved on, and when the interface lies directly between these atoms, to obtain mirror symmetry, they will have displaced exactly half of their final value. This is can be perhaps locally appreciated in some of the simulations by \citet{Serra1988,Serra1991}.

While shuffle displacements are not always along rational planes when $K_1$ is corrugated, net shuffles vectors are always found to be along rational directions and have rational magnitudes, regardless of the Burgers vectors and direction of shuffle displacements. This is a function of the symmetry involved in having shear and mirror transformations which accomplish the same strain tensor. The symmetry of the shuffle mechanisms is clearly seen in that all of their shear plane components lie precisely on shear invariants. This is due to the fact that there is no geometric difference between the mechanism of transforming from parent to twin, and transforming from twin to parent, so that the net shuffles must be invariant under the action of the simple shear.

When there are shear plane displacement components which do not lie in the $\boldsymbol{\eta}_1$ direction, the net shuffle vectors will depend on the order of \emph{Shear}~$\cup$~\emph{Shuffle}. The order, however, must make a difference only in terms of coordinates, because the parent is the twin of the twin, and thus, the \emph{Shear}~$\cup$~\emph{Shuffle} mechanism must be symmetric with a \emph{Shuffle}~$\cup$~\emph{Shear} mechanism. This corresponds to the shuffle acting along $\boldsymbol{\eta}_{2/n}$ and then shear being symmetric with the opposite order and shuffle on $\boldsymbol{\acute{\eta}}_{2/n}$. Applying first shuffle and then shear is more convenient when using Miller indices based on the parent lattice because the shuffle is on $\boldsymbol{\eta}_{2/n}$ and therefore rational in the parent lattice.

\section{Conclusions}

This paper identified a straightforward deterministic theory for analytically deriving both shuffle and shear candidates for any given compound twin mode in HCP metals. The formulations enabled identification of which planes can undergo only shear and which ones must also be the subject of shuffles for any overlying dislocation with a higher step. Application of the theory to \hkl{11-22} and \hkl{10-12} twinning modes revealed that there is an exact repetition of the minimum deviation from the stacking sequence, which corresponds to the smallest possible shear. This enabled identification of a $\mathbf{b}_7$ dislocation for the \hkl{11-22} case, not previously reported in the literature. Also, a twinning dislocation with a step height equal to multiple interplanar spacings does not necessarily require shuffles within intermediate planes to operate in the twinning direction. This is interesting for the case of  $\mathbf{b}_4$ dislocation,  $\mathbf{x}$ shuffles, in the second plane. The analytical expression of net shuffles is useful to calculate the specific energy, excluding shear, required by shuffles during deformation twinning. The net shuffle should be correlated to hydrostatic pressure while plastic shear strain by twinning is deviatoric by definition. This sheds light on the observed sensitivity of twin nucleation to stress concentrations, triaxiality, and temperature.

\section*{Acknowledgments}

The authors would like to recognize the National Science Foundation which supported this work under the DMREF (Designing Materials to Revolutionize and Engineer our Future) program with the award number: CMMI-1235009.  MAT would like to acknowledge support from the U.S. Army Research Laboratory (ARL) administered by the Oak Ridge Institute for Science and Education through an interagency agreement between the U.S. Department of Energy and ARL.

\bibliographystyle{model3-num-names_no_titles}
\bibliography{Master}

\end{document}